\documentclass[nonacm]{acmart}

\AtBeginDocument{%
  }

\setcopyright{acmcopyright}
\copyrightyear{2018}
\acmYear{2018}
\acmDOI{XXXXXXX.XXXXXXX}

\acmJournal{POMACS}
\acmVolume{37}
\acmNumber{4}
\acmArticle{111}
\acmMonth{8}

\usepackage{graphicx}
\usepackage{xcolor}
\usepackage{amsmath}
\usepackage{balance}
\usepackage{comment}
\usepackage{multirow}

\newcommand{\mitre}{MITRE}
\newcommand{\attack}{ATT\&CK}

\begin{document}

\title{MITRE ATT\&CK: State of the Art and Way Forward}

\author{Bader Al-Sada}
\authornotemark[1]
\email{balsada@hbku.edu.qa}
\author{Alireza Sadighian}
\email{asadighian@hbku.edu.qa}
\affiliation{%
  \institution{Division of Information and Computing Technology, College of Science and Engineering, Hamad Bin Khalifa University, Qatar Foundation}
  \city{Doha}
  \country{Qatar}
}
\authornote{Both authors contributed equally to this research.}

\author{Gabriele Oligeri}
\email{goligeri@hbku.edu.qa}
\affiliation{%
  \institution{Division of Information and Computing Technology, College of Science and Engineering, Hamad Bin Khalifa University, Qatar Foundation}
  \city{Doha}
  \country{Qatar}
}

\begin{abstract}
  MITRE ATT\&CK is a comprehensive framework of adversary tactics, techniques and procedures based on real-world observations. It has been used as a foundation for threat modelling in different sectors, such as government, academia and industry. To the best of our knowledge, no previous work has been devoted to the comprehensive collection, study and investigation of the current state of the art leveraging the \mitre\ \attack\ framework. We select and inspect more than fifty major research contributions, while conducting a detailed analysis of their methodology and objectives in relation to the \mitre\ \attack\ framework. We provide a categorization of   the identified papers according to different criteria such as use cases, application scenarios, adopted methodologies and the use of additional data. Finally, we discuss open issues and future research directions involving not only the \mitre\ \attack\ framework but also the fields of risk analysis and cyber-threat intelligence at large.
\end{abstract}



%\ccsdesc[500]{Computer systems organization~Embedded systems}


\received{20 February 2007}
\received[revised]{12 March 2009}
\received[accepted]{5 June 2009}

\maketitle

\section{Introduction}
\label{sec:introduction}

Inter-connectivity and smart technologies are becoming more and more popular in our daily life. The paradigms of fog, edge and cloud computing enable the end-user to interact with an increasing number of remote services. The global trend of ``get everything connected" is affecting the industry (Industry 4.0), cities (smart cities), houses (smart homes), etc. Nevertheless, the high density of online services, sensors, and actuators is increasing the likelihood of cyber-security attacks given the amount of known (and unknown, i.e, zero-day) vulnerabilities that are spread in all the on-line services. The perimeter to be monitored, a.k.a., surface of attack, by each organization is becoming too large to guarantee a zero-risk of penetration---eventually resulting on the acceptance of a minimum risk. Given the growing complexity of ICT infrastructures, and the associated cyber-security countermeasures, modern cyber-security attacks are constituted by a sequence of steps that usually initiate from the weakest point in the attack surface (service security mis-configuration), and then, they move towards their final objective, e.g., data exfiltration, denial of service, subversion, etc. Indeed, cyber-security attacks are today's constituted by a sequence of malicious activities performed over long period of time---this is to guarantee stealthiness during the lateral spreading of the attack. The sequence of malicious operations, a.k.a., adversary behaviour, is a fingerprint of the adversary itself. Indeed, the adversary aims at optimizing his workflow re-using the same tools, scripts and the overall methodology to gain access to a system. This pattern makes each adversary different and (at certain extent) identifiable. The \mitre\ \attack\ framework is a database of attacks, allowing a cross-match between adversarial objectives, behaviour and attack sources. Firstly, \mitre\ \attack\ enables a detailed description of the threat landscape associated with a specific business---this being extremely important to allow the organization to prioritize the deployment of the assets' controls. Moreover, for each attack in the database, \mitre\ \attack\ provides the chain of malicious activities from the initial access to the final attack's objective. This steps represent the adversarial behaviour, which in turn, they can be considered as the fingerprint of the attack allowing the identification of the adversary. Finally, \mitre\ \attack\ potentially allows the forecasting of the subsequent actions when the attack is detected as ongoing in the system---this allowing the localization and mitigation of the attack itself. This survey focuses on the \mitre\ \attack\ knowledge-base applications and its enhancement and extension methods. To this aim, we identified a comprehensive list of papers that consider the \mitre\ \attack\ framework in various ways. Next, we analyzed this list from different perspectives, such as adopted techniques, application scenario, type of their inputs, evaluation techniques, etc., and present our findings and insights. 

{\bf Contributions.} The contributions of this paper are manyfold. Firstly, we provide a comprehensive list of state of the art research papers related to the \mitre\ \attack\ framework. We summarize each research paper describing its methodology, evaluation technique and future directions. Subsequently, we categorize the list of papers into 4 main categories: (i) Behavioral analytic, (ii) Adversary emulation / Red teaming, (iii) Defensive gap assessment, and finally, (iv) CTI enrichment.
We describe our findings and insights according to different categories, such as application scenarios, input types, adopted techniques, and input datasets. Finally, we present the limitations and the existing challenges in the current state of the art while providing some future directions. 

\textbf{Roadmap}. The rest of this paper is organized as follows. Section~\ref{sec:mitre} is dedicated to the introduction of the background knowledge about the \mitre\ \attack\ framework. Section~\ref{sec:comparison} compares the \mitre\ \attack\ with other frameworks, while Section~\ref{sec:use_cases_mapping} introduces the list of the selected paper and a preliminary break-down into use cases. Section~\ref{sec:summary-analysis} wraps up on the key-findings of our survey, while Section~\ref{sec:open-issues-and-research-directions} highlights open issues and future research directions. Finally, Section~\ref{sec:conclusion} draws some conclusion remarks.

\section{Background on MITRE ATT\&CK framework}
\label{sec:mitre}
\begin{figure}
    \centering
    \includegraphics[width=0.7\columnwidth]{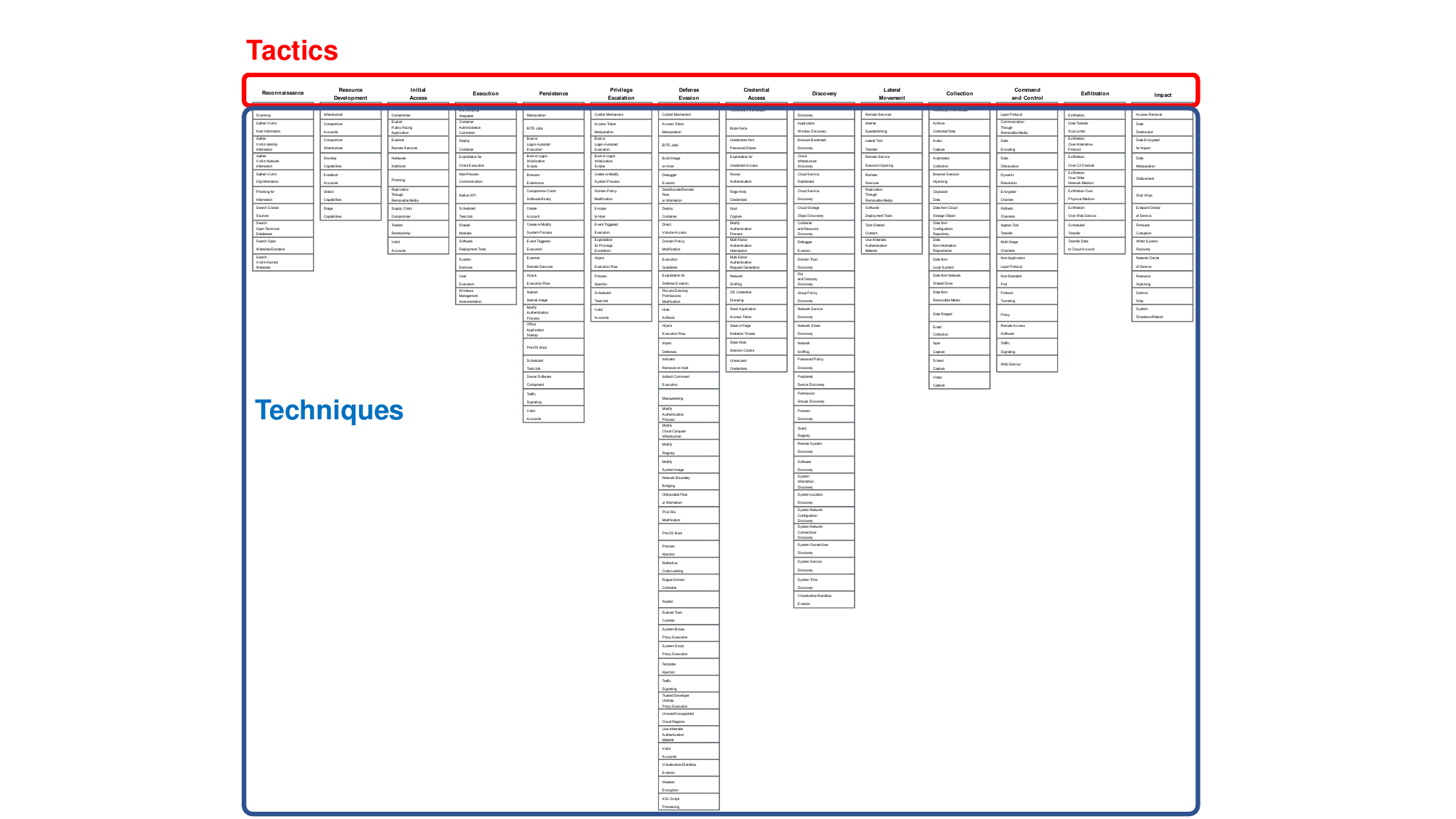}
    \caption{MITRE ATT\&CK matrix: tactics and techniques.}
    \label{fig:attackmatrix}
\end{figure}

MITRE ATT\&CK stands for MITRE Adversarial Tactics, Techniques, and Common Knowledge (\attack). It has been created in 2013 as a result of MITRE's Fort Meade Experiment (FMX) where a group of researchers emulated a scenario constituted by adversaries and defenders in an effort to improve the forensic analysis of an attack through the behavioral analysis of the adversary. The \mitre\ \attack\ framework is a database of malicious activities from real world-observations. Such activities are classified as belonging to \emph{groups} or Advanced Persistent Threat (APTs), thus an APT is characterized by a sequence of malicious activities---this one constituting the behavior of the adversary. The malicious activity is deployed by means of a specific \emph{procedure} which belongs to a \emph{technique} being the mean the adversary adopts to achieve its goal. Finally, techniques are grouped into \emph{tactics}, i.e., the short-term objective of the adversary. Therefore, an APT achieves its objective by implementing a sequence of malicious activities that can be modelled as a sequence of Tactics, Techniques, and Procedures (TTPs).  
Figure~\ref{fig:attackmatrix} shows the typical layout of the \attack\ matrix, where the tactics and the techniques are organized by columns and rows, respectively. Tactics are organized in a logical way (from left to right) in order to highlight the possible consequential phases of an attack.

As a simple example, we consider the well-known \emph{wannacry} ransomware, classified as {\em software} in the \mitre\ framework, i.e., S0366, involving 8 tactics, i.e., Execution, Persistence, Privilege Escalation, Defense Evasion, Discovery, Lateral Movement, Command and Control, and Impact, and 16 techniques, i.e., T1047, T1543, T1222, T1564, T1083, T1120, T1018, T1016, T1210, T1570, T1563, T1573, T1090, T0866,  T1486, T1490, and T1489, as described in the following:

\begin{itemize}
    \item {\em Execution.}
    \begin{itemize}
        \item T1047. Windows Management Instrumentation
    \end{itemize}
    \item {\em Persistence}
    \begin{itemize}
        \item T1543. Create or Modify System Process
    \end{itemize}
    \item {\em Privilege escalation}
    \begin{itemize}
        \item T1543. Create or Modify System Process
    \end{itemize}
    \item {\em Defense evasion}
    \begin{itemize}
        \item T1222. File and Directory Permissions Modification
        \item T1564. Hide Artifacts
    \end{itemize}
    \item {\em Discovery}
    \begin{itemize}
        \item T1083. File and Directory Discovery
        \item T1120. Peripheral Device Discovery
        \item T1018. Remote System Discovery
        \item T1016. System Network Configuration Discovery
    \end{itemize}
    \item {\em Lateral movement}
    \begin{itemize}
        \item T0866. Exploitation of Remote Services
        \item T1570. Lateral Tool Transfer
        \item T1563. Remote Service Session Hijacking
    \end{itemize}
    \item {\em Command and Control}
    \begin{itemize}
        \item T1573. Encrypted Channel
        \item T1090. Proxy
    \end{itemize}
    \item {\em Impact.}
    \begin{itemize}
        \item T1486. Data Encrypted for Impact.
        \item T1490. Inhibit System Recovery.
        \item T1489. Service Stop.
    \end{itemize}
\end{itemize}

\begin{figure}
    \centering
    \includegraphics[width=0.7\columnwidth]{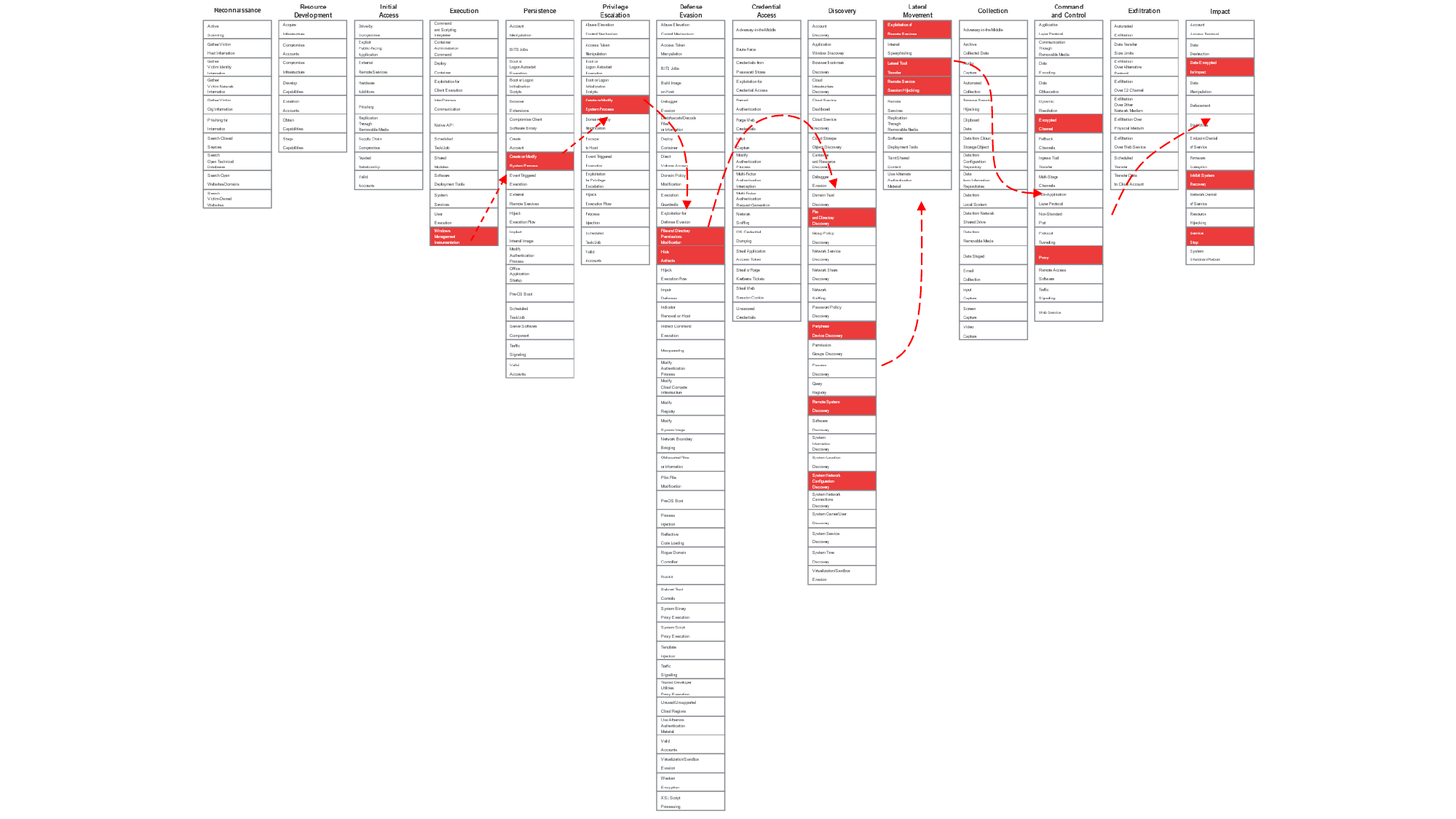}
    \caption{WannaCry behaviour as a sequence of techniques.}
    \label{fig:wannacry_paths}
\end{figure}

Figure~\ref{fig:wannacry_paths} shows the behaviour of WannaCry malware in the \attack\ matrix and the sequence of observed tactics and techniques. In particular, the sequence of the techniques represents the behaviour of the malware also known as fingerprint.

The \attack\ matrix also refers to {\em sub-techniques}, i.e., more specific techniques adopted by the adversary to achieve its goals. As an example, still related to \emph{wannacry} malware, we may consider T1573 (Encrypted Channel), which is characterized by 2 sub-techniques: T1573.001 (Symmetric Cryptography) and T1573.002 (Asymmetric Cryptography)---the later one adopted by \emph{wannacry}. Each technique (or sub-technique) can be implemented in different ways and a list of implementations are considered within the framework, i.e., the procedures associate with sub-technique T1573.002 are 55. The framework also links each considered malicious software, such as \emph{wannacry} - S0366, to the {\em groups} well-known to have adopted it before, i.e., G0032 - Lazarous Group. The \attack\ framework also refers to well-known techniques to \emph{detect} and \emph{mitigate} each of the technique and sub-technique previously discussed. As a final example, M1031 (Network Intrusion Prevention) and M1020 (SSL/TLS Inspection) are suggested as mitigation to sub-technique T1573.002, while as for detection, DS0029 (Network Traffic) is suggested.
  
\subsection{Framework evolution over time}
\label{sec:framework_evolution}
%% MITRE ATTACK over time
% https://d3security.com/blog/see-the-evolution-of-the-mitre-attack-framework-from-2015-to-now/
% https://www.techtarget.com/searchsecurity/news/252491169/Mitre-ATTCK-How-it-has-evolved-and-grown
% https://medium.com/mitre-attack/whats-next-for-mitre-att-ck-7f9506ea1fee

% the latest version has 14 Tactics, 188 Techniques, 379 Sub-techniques, 129 Groups, and 637 Pieces of Software

\begin{figure}
    \centering
    \includegraphics[width=0.7\columnwidth]{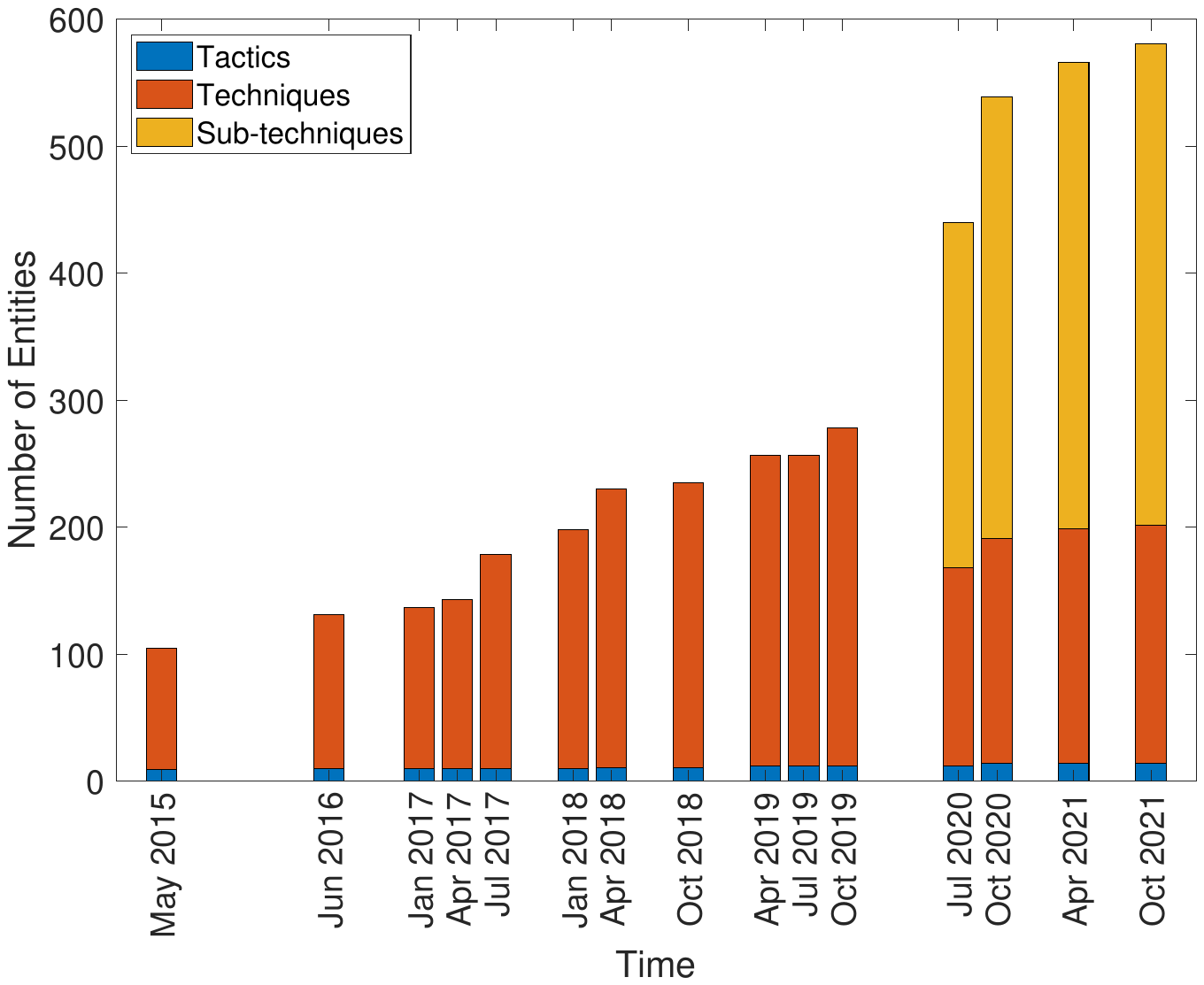}
    \caption{\mitre\ \attack\ framework evolution over time in terms of number of Tactics, Techniques and Sub-Techniques.}
    \label{fig:mitre_evolution}
\end{figure}

The \attack\ matrix is updated (on average) every 6 months with major changes since 2015 (launched as a wiki) reaching the version 11 at the date of the writing of this manuscript. Figure~\ref{fig:mitre_evolution} shows the evolution of the \mitre\ \attack\ framework over the time including the counting of tactics, techniques and sub-techniques~\cite{attack_updates}. A worth-mentioning major change has been introduced with PRE-\attack\ in 2017 involving tactics and techniques for describing the adversarial behaviour before the actual attack deployment; subsequently, PRE-\attack\ has been integrated in the \attack\ framework as the columns {\em Reconnaissance} and {\em Resource Development}. At the same time (July 2017), Linux and macOS have be been introduced in the framework. In October 2019, \attack\ for cloud was added to the framework covering the cloud-based Infrastructure as a Service (IaaS). Finally, \attack\ for Industrial Control Systems (ICSs) has been introduced in the 2020.

\subsection{Technology domains}
\label{sec:technology_domains}

The \mitre\ attack\ framework involves three technology domains: {\em Enterprise}, {\em Mobile}, and {\em ICS}. The three domains involve different tactics and techniques---although there are some minor overlaps. Indeed, the three domains are very different each others, thus involving different adversarial behaviour. The attack surfaces are different as well as the adversary objectives, thus requiring ad-hoc tactics, techniques and procedures.

{\em Enterprise.} This domain involves 191 techniques and 385 sub-techniques including several platform such as Windows, Linux, macOS, etc. This is the richest domain due to its legacy involving observations of attacks from a technology domain that is physiologically broader than the others. Indeed, the enterprise domain involves the platform Windows, macOS, Linux, Cloud, Network and Containers, while being the only one to include specific techniques (within 2 tactics) for the analysis of the preparatory adversarial activities (PRE-\attack). Indeed, worth-noting entry points in the \attack\ matrix are Search Open Websites/Domains and Search Victim-Owned Websites (Reconnaissance)---typical behaviour for social engineering, while typical adversary objectives are Data Destruction, Disk Wipe and Service Stop (Impact)---techniques that commonly implemented by adversaries in enterprise scenarios.

{\em Mobile.} The mobile domain takes into account 66 techniques and 41 sub-techniques. The mobile domain has the main objective to perform a behavioral analysis of attacks involving smart-phones and related devices. The increasing adoption of smart-phones for accessing and sharing personal and corporate data, e.g., Bring Your Own Device model (BYOD), significantly expand the attack surface while increasing the number of the associated threats. Indeed, countless attacks have been proposed against smart-phones, and those attacks are becoming more and more targeted to the user's organization, given the entry points offered by the smart-phones vulnerabilities. The mobile \attack\ matrix addresses this scenario considering specific attack behaviour with typical entry points of smart-phones, e.g., Lockscreen Bypass and Supply Chain Compromise (Initial Access), while aiming at targets that can involve more than the user's data, e.g., Input Injection and Data Encrypted for Impact (Impact).

{\em Industrial Control Systems (ICSs).} The ICS domain is represented by 78 techniques with no sub-techniques. Modern cyber-physical systems are experience the convergence of Information Technology (IT) and Operational Technology (OT) throughout the new models of fog and edge computing. Indeed, the OT infrastructure is now connected to the network with processing capabilities---being able to implement AI techniques in some cases, thus enabling the full integration of data-centric computing (IT) with monitoring and sensing from the physical world (OT). While the increasing integration of OT and IT enhance the effectiveness and efficacy of the industrial processes, it represents an unprecedented opportunity for adversaries than can now jump from one world to the other, thus increasing their chances to affect the physical environment by initiating attacks in the cyber-space. The ICS \attack\ matrix captures this IT-OT relation by considering attack behaviour that initiate from the IT world, e.g., Spearfishing attachment, Wireless Compromise, Internet Accessible Device (Initial access), ending up with a physical impact in the OT world, e.g., Damage to Property, Denial of Control, and Loss of Safety (Impact).

\section{Comparison with other frameworks}
\label{sec:comparison}

There are mainly 2 other frameworks that can support the risk analysis activity and they can be compared with the \mitre\ \attack\ framework: (i) the Lockheed Martin's Cyber Kill Chain (CKC) and (ii) the Microsoft STRIDE framework (STRIDE). Two other minor models are also considered: the Unified Kill Chain (UKC) and the Diamond model (DM).

{\bf Lockheed Martin's Cyber Kill Chain.} This framework involves 7 stages, i.e., Reconnaissance, Weaponization, Delivery, Exploitation, Installation, Command and Control, and Actions on Objectives. The idea behind the framework is rooted on the fact that all the cybersecurity attacks can be either represented or mapped in terms of a sequence of temporal activities belonging to the aforementioned stages. It is strictly important that the temporal sequence is respected: the success of the attack is tight to the temporal sequence of the adversary's actions. Indeed, this makes the attack (or the perception of it) much weaker: any action that breaks the chain (sequence of malicious activities) prevents the adversary to achieve its goal. 

{\bf Microsoft STRIDE.} This framework has been proposed and developed by Praerit Garg and Loren Kohnfelder at Microsoft~\cite{stride}. The model leverages 6 threat categories: Spoofing, Tampering with data, Repudiation, Information disclosure, Denial of service and Elevation of privilege. The main idea is to identify a specific service from the corporate architecture, and then, adopt the 6 categories to identify all the possible threats, and finally, iterating the methodology for each service in the organization. This framework allows to identify vulnerabilities, and the associated threats, while been driven by the STRIDE categories.
\\
\\
Although the three frameworks have been analyzed and dissected by several blogs, they cannot be really compared. Indeed, \attack\ is mainly a database of malicious activities, CKC is an attack model based on sequence of causal activities, while STRIDE is a category-based threat identification process. The objectives and the scopes of the frameworks are different: \attack\ aims at supporting cybersecurity experts with a wide range of previously identified attacks and the associated mitigation actions, while CKC is a general attack model that can be exploited to identify the most effective response strategies, and finally, STRIDE aims at supporting the threat identification process in the corporate environment. 

While serving different purposes, their integration and inter-operation is still at the early stage. For example, STRIDE might be used to identify a threat associated with one or more services in the organization, CKC might support the modelling of the attack associated with the threat while highlighting potential mitigation, and finally, \mitre\ \attack\ can be used to identify similar attacks, the associated groups, the activities (techniques) deployed to achieve the adversarial goal, and finally, identify the countermeasures. 

\begin{figure}
    \centering
    \includegraphics[width=0.8\columnwidth]{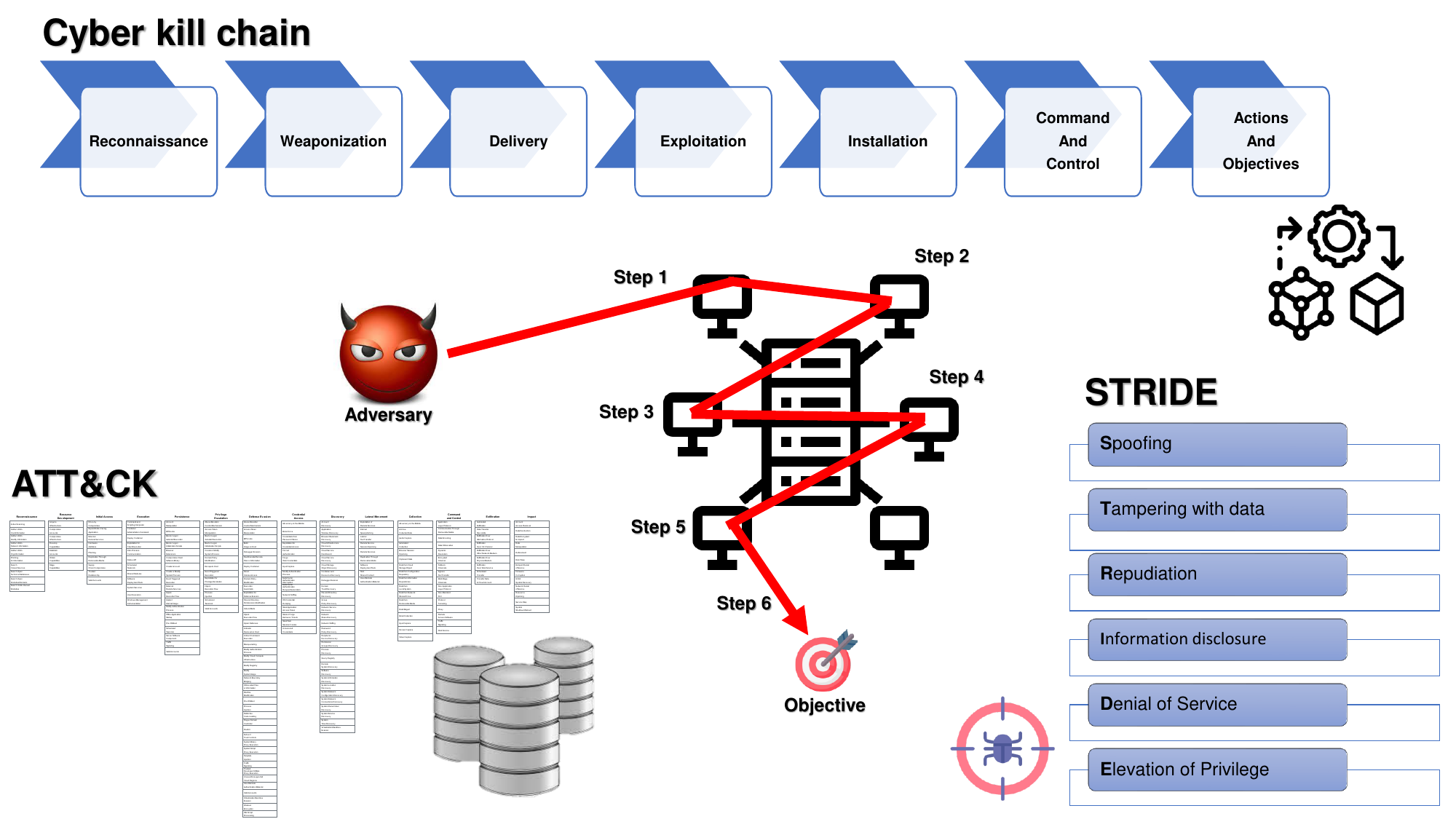}
    \caption{\attack, Cyber Kill Chain and {STRIDE}: comparison and application scenarios.} 
    \label{fig:comparison}
\end{figure}

Figure~\ref{fig:comparison} shows a diagram (center of the figure) associated with a generic cybersecurity attack involving several steps (from Step 1 to Step 6) affecting the ICT infrastructure before reaching the final adversary objective. The STRIDE framework can support the analysis and early identification of vulnerabilities associated with the entry point (Step 1), each of the services exploited by the adversary to move throughout the ICT infrastructure (Step 2 TO 5), and finally, the vulnerability associated with the target objective (Step 6). Conversely, CKC can be effectively adopted to create a model of the consecutive and causal events that allows the adversary to reach its objective (Step 6) from early stages (Step 1)---it is worth highlighting that a proper modelling can help the security risk assessment team to break the attack at each step by deploying the correct countermeasure thus preventing the adversary to move forward and reaching its objective.

The {\bf Unified Kill Chain} (UKC)~\cite{pols2017unified} has been proposed by Paul Pols in 2017 as the result of his Master Thesis. UKC both extends and improves the CKC by considering threats that go beyond malware and the corporate perimeter. Nevertheless, UKC is still based on a (more extended) sequence of adversary actions being based on the concept of {\em kill chain}.

Finally, we report the {\bf Diamond model} (DM)~\cite{caltagirone2013diamond} mainly considered in the field of intrusion analysis. DM involves 4 entities, i.e., the adversary, the capability, the infrastructure, and the victim. The analysis of the intrusion is performed by considering a description of the ``adversary", its ``capabilities", how it interacts with the target ``infrastructure" in order to achieve its objectives against the ``victim".

\section{In-depth Analysis and Use Cases Mapping}
\label{sec:use_cases_mapping}

In this section, we introduce the list of the papers taken into account in this survey. The resulting list has been obtained by searching for the keywords ``\mitre" and ``\attack" in Google Scholar. We only consider the contributions where the \mitre\ \attack\ framework is a strict requirement for achieving the objectives of the paper, while we discarded all the other ones.

Table~\ref{table:main-list-table1} and~\ref{table:main-list-table2} shows the list of the considered papers, where we report the title of the paper, the year of the publication, the authors' country, and finally, the number of citations collected up to the writing of this survey.

\begin{table}
\footnotesize
\caption{Our reference list of papers (1/2).}
\centering
\begin{tabular}{clclc}
{\bf Reference}       & {\bf Title} & {\bf Year} & {\bf Country} & {\bf N. Citations} \\
\hline
\hline
\cite{milajerdi2019holmes}        & \begin{tabular}[c]{@{}l@{}} HOLMES: Real-time APT Detection through Correlation of Suspicious Information\\ Flows \end{tabular} & 2019 & US & 229 \\
\cite{mavroeidis2017cyber}        & \begin{tabular}[c]{@{}l@{}} Cyber Threat Intelligence Model: An Evaluation of Taxonomies, Sharing Standards,\\ and Ontologies within Cyber Threat Intelligence \end{tabular} & 2017 & NO & 214 \\
\cite{husari2017ttpdrill}         & \begin{tabular}[c]{@{}l@{}} TTPDrill: Automatic and Accurate Extraction of Threat Actions from Unstructured \\Text of CTI Sources \end{tabular} & 2017 & US & 98 \\
\cite{hassan2020tactical}         & \begin{tabular}[c]{@{}l@{}} Tactical Provenance Analysis for Endpoint Detection and Response Systems \end{tabular} & 2020 & US & 90 \\
\cite{rawan2020}             & \begin{tabular}[c]{@{}l@{}} Learning the associations of \mitre\ \attack\ adversarial techniques \end{tabular} & 2020 & US & 45 \\
\cite{noor2019machine}            & \begin{tabular}[c]{@{}l@{}} A machine learning framework for investigating data breaches based on semantic\\ analysis of adversary’s attack patterns in threat intelligence repositories \end{tabular} & 2019 & US,PK & 44 \\
\cite{xiong2022cyber}             & \begin{tabular}[c]{@{}l@{}} Cyber security threat modeling based on the \mitre\ Enterprise \attack\ Matrix \end{tabular} & 2022 & SE, FR & 41 \\
\cite{stech2016integrating}       & \begin{tabular}[c]{@{}l@{}} Integrating Cyber-D\&D into Adversary Modeling for Active Cyber Defense \end{tabular} & 2016 & US & 33 \\
\cite{applebaum2017analysis}      & \begin{tabular}[c]{@{}l@{}}Analysis of automated adversary emulation techniques \end{tabular} & 2017 & US & 30 \\
\cite{oosthoek2019sok}            & \begin{tabular}[c]{@{}l@{}} Sok: Att\&ck techniques and trends in windows malware \end{tabular} & 2019 & NL & 24 \\
\cite{farooq2018optimal}          & \begin{tabular}[c]{@{}l@{}} Optimal Machine Learning Algorithms for Cyber Threat Detection \end{tabular} & 2018 & SA & 23 \\
\cite{georgiadou2021assessing}    & \begin{tabular}[c]{@{}l@{}} Assessing \mitre\ \attack\ Risk Using a Cyber-Security Culture Framework \end{tabular} & 2021 & GR & 21 \\
\cite{tatam2021review}            & \begin{tabular}[c]{@{}l@{}} A review of threat modelling approaches for APT-style attacks \end{tabular} & 2021 & AU & 20 \\
\cite{kwon2020cyber}              & \begin{tabular}[c]{@{}l@{}} Cyber threat dictionary using \mitre\ \attack\ matrix and NIST cybersecurity\\ framework mapping \end{tabular} & 2020 & US & 19 \\
\cite{munaiah2019characterizing}  & \begin{tabular}[c]{@{}l@{}} Characterizing Attacker Behavior in a Cybersecurity Penetration Testing Competition\end{tabular}  & 2019 & US & 17 \\
\cite{straub2020modeling}         & \begin{tabular}[c]{@{}l@{}} Modeling attack, defense and threat trees and the cyber kill chain, \attack\ and \\ STRIDE frameworks as blackboard architecture networks \end{tabular} & 2020 & US & 17 \\
\cite{hacks2020powerlang}         & \begin{tabular}[c]{@{}l@{}} powerLang: a probabilistic attack simulation language for the power domain \end{tabular} & 2020 & SW & 15 \\
\cite{nisioti2021data}            & \begin{tabular}[c]{@{}l@{}} Data-Driven Decision Support for Optimizing Cyber Forensic Investigations \end{tabular} & 2021 & US, UK & 15\\
\cite{legoy2020automated}         & \begin{tabular}[c]{@{}l@{}} Automated Retrieval of \attack\ Tactics and Techniques for Cyber Threat Reports \end{tabular} & 2020 & NL, DE & 15 \\
\cite{husari2019learning}         & \begin{tabular}[c]{@{}l@{}} Learning APT Chains from Cyber Threat Intelligence \end{tabular} & 2019 & US & 14 \\
\cite{zilberman2020sok}           & \begin{tabular}[c]{@{}l@{}} SoK: A survey of open-source threat emulators \end{tabular} & 2020 & IL & 14 \\
\cite{maymi2017towards}           & \begin{tabular}[c]{@{}l@{}} Towards a Definition of Cyberspace Tactics, Techniques and Procedures \end{tabular} & 2017 & US & 14 \\
\cite{parmar2019use}              & \begin{tabular}[c]{@{}l@{}}  On the Use of Cyber Threat Intelligence (CTI) in Support of Developing the Commander's \\ Understanding of the Adversary \end{tabular} & 2019 & NATO & 11 \\
\cite{choi2021probabilistic}      & \begin{tabular}[c]{@{}l@{}} Probabilistic Attack Sequence Generation and Execution Based on \mitre\ \attack\ \\ for ICS Datasets \end{tabular} & 2021 & KR & 10 \\
\cite{berady2021ttp}              & \begin{tabular}[c]{@{}l@{}} From TTP to IoC: Advanced Persistent Graphs for Threat Hunting \end{tabular} & 2021 & FR & 9 \\
\cite{hasan2019artificial}        & \begin{tabular}[c]{@{}l@{}} Artificial Intelligence empowered Cyber Threat Detection and Protection for \\ Power Utilities \end{tabular} & 2019 & US & 9 \\
\cite{elitzur2019attack}          & \begin{tabular}[c]{@{}l@{}} Attack Hypothesis Generation \end{tabular} & 2019 & IL & 8 \\
\cite{kuppa2021linking}           & \begin{tabular}[c]{@{}l@{}} Linking CVE’s to \mitre\ \attack\ Techniques \end{tabular} & 2021 & IE, FR & 8  \\
\cite{choi2020expansion}          & \begin{tabular}[c]{@{}l@{}} Expansion of ICS Testbed for Security Validation based on \mitre\ \attack\ Techniques \end{tabular} & 2020 & KR & 8  \\
\cite{kim2021automatically}       & \begin{tabular}[c]{@{}l@{}} Automatically Attributing Mobile Threat Actors by Vectorized \attack\ Matrix \\ and Paired Indicator \end{tabular} & 2021 & KR & 7 \\
\cite{golushko2020application}    & \begin{tabular}[c]{@{}l@{}} Application of Advanced Persistent Threat Actors' Techniques for Evaluating \\ Defensive Countermeasures \end{tabular} & 2020 & RU & 6 \\
\cite{karuna2021automating}       & \begin{tabular}[c]{@{}l@{}} Automating Cyber Threat Hunting Using NLP, Automated Query Generation, and \\ Genetic Perturbation \end{tabular} & 2021 & US & 4 \\ 
\cite{wu2020grouptracer}          & \begin{tabular}[c]{@{}l@{}} GroupTracer: automatic attacker TTP profile extraction and group cluster in \\ Internet of things \end{tabular} & 2020 & CH & 4 \\ 
\hline
\end{tabular}
\label{table:main-list-table1}
\end{table}

\begin{table}
\footnotesize
\caption{Our reference list of papers (2/2).}
\centering
\begin{tabular}{clclc}
{\bf Reference}       & {\bf Title} & {\bf Year} & {\bf Country} & {\bf N. Citations} \\
\hline
\hline
\cite{ampel2021linking}           & \begin{tabular}[c]{@{}l@{}} Linking Common Vulnerabilities and Exposures to the \mitre\ \attack\ Framework: \\ A Self-Distillation Approach \end{tabular} & 2021 & US & 4 \\
\cite{jadidi2021threat}           & \begin{tabular}[c]{@{}l@{}} A Threat Hunting Framework for Industrial Control Systems \end{tabular} & 2021 & AU & 4 \\
\cite{arshad2021attack}           & \begin{tabular}[c]{@{}l@{}} Attack Specification Language: Domain Specific Language for Dynamic Training\\ in Cyber Range \end{tabular} & 2021 & PK & 3 \\
\cite{bertino2021services}        & \begin{tabular}[c]{@{}l@{}} Services for zero trust architectures-a research roadmap \end{tabular} & 2021 & US & 3 \\
\cite{toker2021mitre}             & \begin{tabular}[c]{@{}l@{}} \mitre\ ICS Attack Simulation and Detection on EtherCAT Based Drinking Water System \end{tabular} & 2021 & TR & 3 \\
\cite{bertino2021services}        & \begin{tabular}[c]{@{}l@{}} Services for zero trust architectures-a research roadmap \end{tabular} & 2021 & US & 3 \\
\cite{mendsaikhan2020automatic}   & \begin{tabular}[c]{@{}l@{}} Automatic Mapping of Vulnerability Information to Adversary Techniques \end{tabular} & 2020 & JP & 3 \\
\cite{manocha2021security}        & \begin{tabular}[c]{@{}l@{}} Security Assessment Rating Framework for Enterprises using \mitre\ \attack\ Matrix \end{tabular} & 2021 & IN & 2  \\
\cite{wang2020clustering}         & \begin{tabular}[c]{@{}l@{}} Clustering using a similarity measure approach based on semantic analysis of \\ adversary behaviors \end{tabular} & 2020 & CN & 2 \\
\cite{takey2021real}              & \begin{tabular}[c]{@{}l@{}} Real Time early Multi Stage Attack Detection \end{tabular} & 2021 & IN & 2 \\
\cite{charan2021dmapt}            & \begin{tabular}[c]{@{}l@{}} DMAPT: Study of Data Mining and Machine Learning Techniques in \\Advanced Persistent Threat Attribution and Detection \end{tabular} & 2021 & IN & 2 \\
\cite{wu2021price}                & \begin{tabular}[c]{@{}l@{}} Price TAG: Towards Semi-Automatically Discovery Tactics, Techniques and Procedures of \\ E-Commerce Cyber Threat Intelligence \end{tabular} & 2021 & CN & 2 \\
\cite{you2022tim}                 & \begin{tabular}[c]{@{}l@{}} TIM: threat context-enhanced TTP intelligence mining on unstructured threat data \end{tabular} & 2022 & CH & 1 \\
\cite{huang2021open}              & \begin{tabular}[c]{@{}l@{}} Open Source Intelligence for Malicious Behavior Discovery and Interpretation \end{tabular} & 2021 & TW & 2 \\
\cite{jo2022cyberattack}          & \begin{tabular}[c]{@{}l@{}} Cyberattack Models for Ship Equipment Based on the MITRE ATT\&CK Framework \end{tabular} & 2022 & KR & 1\\
\cite{mashima2022mitre}           & \begin{tabular}[c]{@{}l@{}} \mitre\ \attack\ Based Evaluation on In-Network Deception Technology for \\ Modernized Electrical Substation Systems \end{tabular} & 2022 & SG & 1 \\
\cite{ahmed2022mitre}             & \begin{tabular}[c]{@{}l@{}} \mitre\ \attack -driven Cyber Risk Assessment \end{tabular} & 2022 & UK, GR & 1 \\
\cite{nisioti2021game}            & \begin{tabular}[c]{@{}l@{}} Game-Theoretic decision support for cyber forensic investigations \end{tabular} & 2021 & UK & 1 \\
\cite{pell2021towards}            & \begin{tabular}[c]{@{}l@{}} Towards Dynamic Threat Modelling in 5G Core Networks Based on \mitre\ \attack\ \end{tabular} & 2021 & UK & 1 \\
\cite{al2022fingerprint}                & \begin{tabular}[c]{@{}l@{}} Fingerprint for Mobile-Sensor APT Detection Framework (FORMAP) Based on \\ Tactics Techniques and Procedures (TTP) and \mitre \end{tabular} & 2022 & MY & 1 \\
\cite{liu2022threat}              & \begin{tabular}[c]{@{}l@{}} Threat intelligence \attack\ extraction based on the attention transformer \\hierarchical recurrent neural network \end{tabular} & 2022 & CH & 0 \\
\cite{chierzi2021evolution}       & \begin{tabular}[c]{@{}l@{}} Evolution of IoT Linux Malware: A \mitre\ \attack\ TTP Based Approach \end{tabular} & 2021 & IT & 0 \\
\cite{fairbanks2021identifying}   & \begin{tabular}[c]{@{}l@{}} Identifying \attack\ Tactics in Android Malware Control Flow Graph \\ Through Graph Representation Learning and Interpretability \end{tabular} & 2021 & US & 0\\
\cite{he2021model}                & \begin{tabular}[c]{@{}l@{}} A Model and Method of Information System Security Risk Assessment based on\\ \mitre\ \attack\ \end{tabular} & 2021 & CH & 0 \\
\cite{zych2022enhancing}          & \begin{tabular}[c]{@{}l@{}} Enhancing the STIX Representation of \mitre\ \attack\ for Group Filtering and\\ Technique Prioritization \end{tabular} & 2022 & NO & 0 \\
\cite{ejaz2022visualizing}        & \begin{tabular}[c]{@{}l@{}} Visualizing Interesting Patterns in Cyber Threat Intelligence Using\\ Machine Learning Techniques \end{tabular} & 2022 & PK, SK & 0 \\
\hline
\end{tabular}
\label{table:main-list-table2}
\end{table}

\begin{figure}
    \centering
    \includegraphics[width=0.6\columnwidth]{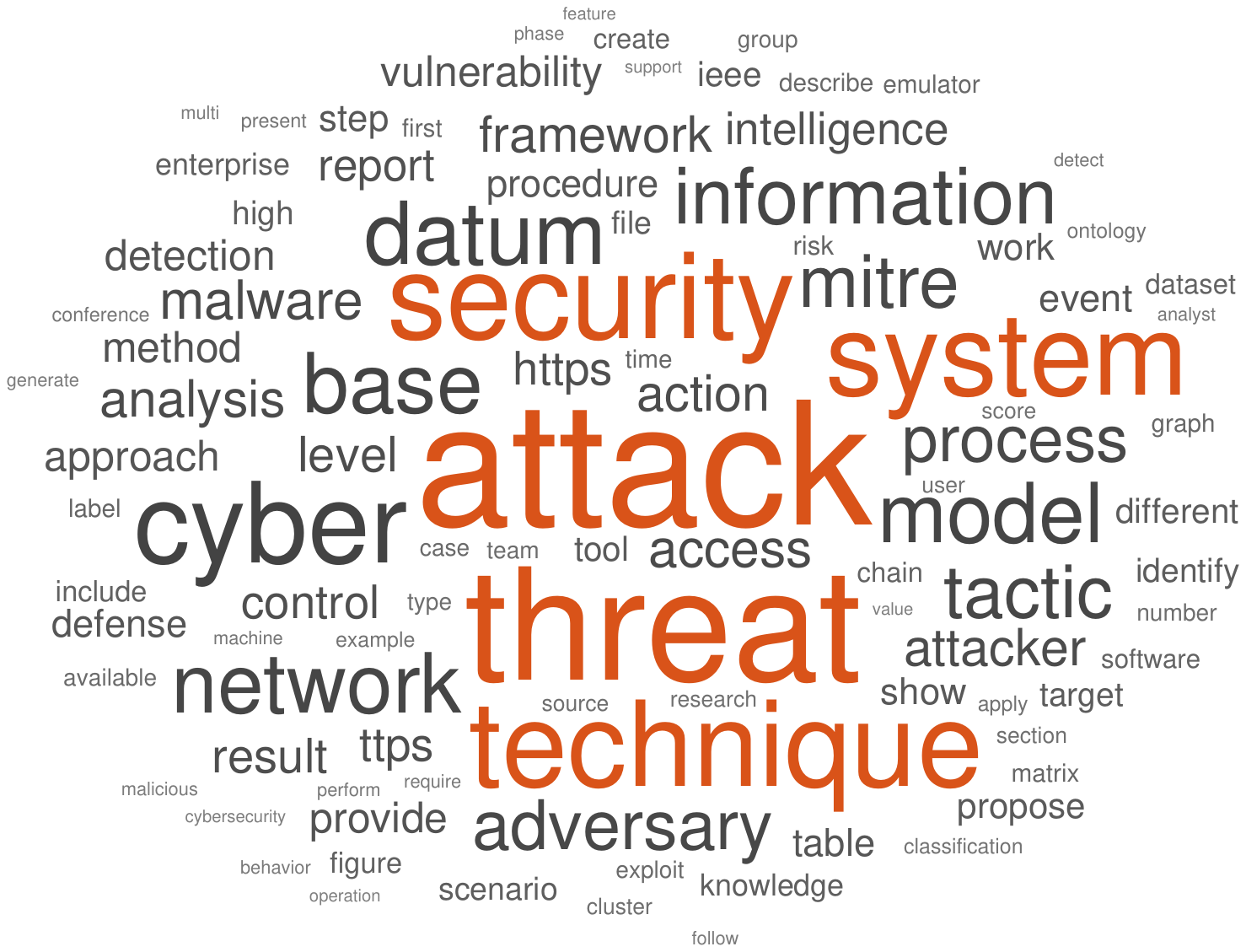}
    \caption{WordCloud computed over the set of papers considered for this survey.} 
    \label{fig:wordcloud}
\end{figure}

Figure~\ref{fig:wordcloud} shows the ``cloud of words" generated from the text extracted from the considered papers. While {\em attack} is the most common word, as expected other common words are {\em threat} and {\em technique} which are core concepts in the framework, while the {\em security} and {\em system} concepts are instantiated depending on the scenario application.

Inspired by the seminal white paper~\cite{strom2018mitre}, we initiate our analysis by categorizing the works as a function of the use cases proposed by~\cite{strom2018mitre}:

\begin{itemize}
    \item[-] Behavioral analytic
    \item[-] Adversary emulation / Red teaming
    \item[-] Defensive gap assessment
    \item[-] CTI enrichment
\end{itemize}

%% Let's add a picture here
Table~\ref{fig:use-case-table} summarizes this categorization, while in the following, we map the papers to each identified category. 

\begin{table}
\footnotesize
    \centering
    \caption{Use-case categories.}
    \begin{tabular}{|l|l|}
    \hline
        \textbf{Use Cases} & \textbf{Papers} \\ \hline\hline
        \textbf{Behavioural analytic} & \begin{tabular}[c]{@{}l@{}}
        \cite{milajerdi2019holmes},
 \cite{hassan2020tactical},
 \cite{rawan2020},
 \cite{noor2019machine},
 \cite{oosthoek2019sok},
 \cite{farooq2018optimal},
 \cite{munaiah2019characterizing},
 \cite{straub2020modeling},
 \cite{hasan2019artificial},
 \cite{karuna2021automating},
 \cite{wu2020grouptracer},
 \cite{jadidi2021threat},
 \cite{toker2021mitre},
 \cite{huang2021open}, \\
 \cite{wang2020clustering},
 \cite{takey2021real},
 \cite{charan2021dmapt},
 \cite{ahmed2022mitre},
 \cite{nisioti2021game},
 \cite{kadhimi2022},
 \cite{chierzi2021evolution},
 \cite{fairbanks2021identifying} \end{tabular}\\ \hline
 
        \textbf{CTI enrichment} & \begin{tabular}[c]{@{}l@{}}
        \cite{mavroeidis2017cyber}, 
 \cite{husari2017ttpdrill}, 
 \cite{stech2016integrating}, 
 \cite{georgiadou2021assessing}, 
 \cite{kwon2020cyber}, 
 \cite{nisioti2021data}, 
 \cite{legoy2020automated}, 
 \cite{husari2019learning}, 
 \cite{maymi2017towards}, 
 \cite{parmar2019use}, 
 \cite{elitzur2019attack}, 
 \cite{kuppa2021linking}, 
 \cite{choi2020expansion}, 
 \cite{kim2021automatically}, \\
 \cite{ampel2021linking}, 
 \cite{mendsaikhan2020automatic}, 
 \cite{wu2021price}, 
 \cite{you2022tim}, 
 \cite{liu2022threat}, 
 \cite{zych2022enhancing}, 
 \cite{ejaz2022visualizing}  \end{tabular}\\ \hline
        \textbf{Adversary Emulation / Red teaming} & \cite{xiong2022cyber}, 
        \cite{applebaum2017analysis}, 
 \cite{tatam2021review}, 
 \cite{hacks2020powerlang},  
 \cite{zilberman2020sok}, 
 \cite{choi2021probabilistic}, 
 \cite{berady2021ttp}, 
 \cite{arshad2021attack}, 
 \cite{pell2021towards} \\ \hline
        \textbf{Defensive Gap Assessment} &  \cite{golushko2020application}, 
 \cite{bertino2021services}, 
 \cite{manocha2021security}, 
 \cite{jo2022cyberattack}, 
 \cite{he2021model} \\ \hline
    \end{tabular}
    \label{fig:use-case-table}
\end{table}

\subsection{Behavioural analytic}

In this section, we consider all the papers addressing the analytical description of the adversary behaviour throughout different methodologies, input and metrics. Indeed, the behaviour of the adversary is a key component for both forensic and forecasting purposes, while the \attack\ framework represents a rich source of information to this goal. In the remainder of this section, we consider all the papers dealing with the analysis of the adversarial behaviour by either considering the data from the \attack\ matrix or from real-world dataset which are mapped to the tactics, techniques and procedures in the \attack\ matrix.

\subsubsection{Learning the Associations of \mitre\ \attack\ Adversarial Techniques, 2020,~\cite{rawan2020}} 

Authors propose a machine learning based approach employing hierarchical clustering in particular to extract the associations of attack techniques from APT and software attack data from the \mitre\ \attack\ framework. This work identifies three key challenges: (i) how important is the establishing an appropriate distance metric for clustering technique associations while maintaining interpretability; (ii) handling multi-dimensional relationships demonstrated by attack techniques; and finally, (iii) Verifying the consistency and importance of the learned hierarchical clustering and techniques associations. For this purpose, authors explore various partitioned clustering methods, e.g., Partion Around Medoids (PAM), K-means, and Fuzzy clustering, and hierarchical methods for clustering attack techniques that take place together in the TTP chain. Nevertheless, 
given the complexity of the adversarial behaviour and the associated techniques constituting the attack, most of the classic clustering algorithms fails to capture usable associations; therefore, based on their results, they choose hierarchical clustering to represent attack patterns. Accordingly, from APT attacks dataset, they are able to infer 37 fine-grain technique associations and from software attack dataset, they are able to infer 61 technique associations. Mutual information has been considered as main criterion for evaluation, and their results show acceptable predictability for APT attacks: 78\% of the fine-grain and 75\% coarse-grain technique associations. 

\subsubsection{Characterizing Attacker Behavior in a Cybersecurity Penetration Testing Competition, 2019,~\cite{munaiah2019characterizing}}
This research proposes an analytical approach to characterise attackers behavior based on a dataset of events collected from the 2018 National Collegiate Penetration Testing Competition (CPTC'18). The main objective is to familiarize developers and administrators with attackers mindsets using tactics and techniques from the \mitre\ \attack\ framework. For this purpose, authors analyze a inter-modal dataset including over 500 million events from six teams of attackers.  The main phases of the proposed approach is evidence gathering by analyzing exploited vulnerabilities and behavior characterization by mapping captured events to the tactics and techniques of the \attack\ framework. The evaluation is performed by applying the identified methodology to the events associated with the different teams.

\subsubsection{Modeling Attack, Defense and Threat Trees and the Cyber Kill Chain, \attack\ and STRIDE Frameworks as Blackboard Architecture Networks, 2020,~\cite{straub2020modeling}} Authors propose the Blackboard Architecture Cyber Command Entity Attack Route (BACCER) that is a generalized approach to model and analyze framework/paradigm-based attacks. While best practices involve the deployment of a single exploit for a single identified target, authors propose to combine rules and facts that demonstrate attack types and corresponding decision making logic with actions: BACCER uses Blackboard Architecture networks~\cite{hayes1985blackboard} to model and analyze attack paradigms.  BACCER's efficiency has been verified by modelling and implementing samples of attack and defense trees. Finally, authors also discussed its efficacy to model other types of frameworks, such as the \mitre\ \attack\ framework.

\subsubsection{Tactical Provenance Analysis for Endpoint Detection and Response Systems, 2020,~\cite{hassan2020tactical}} This research introduces a concept, called Tactical Provenance, that represents casual dependencies between threat alerts generated by Endpoint Detection and Response (EDR) solutions. They encode casual dependencies into Tactical Provenance Graphs (TPG) that are more compressed compared to classical whole-system provenance graphs as of their abstract nature and better visualization capabilities. This helps the security analysts to optimize the inspection process. As the next phase of their research, the authors propose a threat scoring algorithm based on TPGs to assign appropriate scores and rank threat alerts. \mitre\ \attack\ framework is used as a knowledge base for their threat analyzing and processing tasks. The proposed approach is evaluated using the Symantec EDR tool in an enterprise environment. They claim that the achieved results show 87\% overhead reduction in system logs storage. Moreover, the proposed approach enhances the threat detection precision of the Symantec EDR. 

\subsubsection{Sok: Att\&ck techniques and trends in Windows malware, 2019,~\cite{oosthoek2019sok}} This research has conducted an automated analysis and investigation of a sample dataset of 951 Windows malware families mapped and plotted on the \mitre\ \attack\ framework. The authors claim that this research work is the first to use the \mitre\ \attack\ framework to present the results of analyzing a large samples of malware. The malware samples have been collected from Malpedia that is a malware database developed and maintained by Fraunhofer FKIE\footnote{https://malpedia.caad.fkie.fraunhofer.de/}. For malware analysis, the authors use Joe Sandbox Cloud\footnote{https://www.joesandbox.com/} which is a public environment automating dynamic inspection for static and network analysis. Their results show an improvement in the implementation of file-less execution vectors using Windows Management Instrumentation (WMI) and PowerShell, discovery of security software and DLL side-loading for defense evasion. These results indicate how malware developers are empowering their tactics in order to bypass classic and traditional defense strategies.

\subsubsection{A Threat Hunting Framework for Industrial Control Systems, 2021,~\cite{jadidi2021threat}} An ICS Threat Hunting Framework (ICS-THF) for early ICS cyber threat detection is proposed in~\cite{jadidi2021threat}. The proposed framework has three main phases: (i) an initial phase devoted to the detection of any event or external resource triggering the hunting stage, e.g., notifications from third parties and data received from security analysts; (ii) a threat hunting phase that combines and aggregates two source of data, \mitre\ \attack\ framework and Diamond intrusion analysis model, in order to define and build a hunting hypothesis for predicting the future actions of the adversary; and finally, (iii) a Cyber Threat Intelligence (CTI) phase to generate Indicators of Compromises (IoCs), such as TTPs, IPs, etc., to aid organizations learning existing threats. In the proposed framework, CTI is extracted from the actions of identified adversaries. For the evaluation of the proposed approach, they use three scenarios: (i) Black Energy 3 malware, (ii) PLC-Blaster malware, and finally, (iii) SWaT dataset. Based on the obtained results, the authors claim that enterprises can leverage the proposed framework to design and develop customized threat hunting solutions for their own ICS networks.

\subsubsection{MITRE ICS Attack Simulation and Detection on EtherCAT Based Drinking Water System, 2021,~\cite{toker2021mitre}} Authors propose a machine learning based detection approach along with a host based intrusion detection for protecting EtherCAT based drinking water system. In particular, authors  propose a SVM-based detection algorithm for detecting attacks on field devices. Moreover, they use Wazuh HIDS\footnote{https://documentation.wazuh.com/3.13/user-manual/capabilities/anomalies-detection/how-it-works.html} for detecting attacks within the SCADA system control center. Attack vectors have been designed based on the techniques of the \mitre\ ICS \attack\ matrix during their design and development. Finally, they adopted Elasticsearch, Logstash and Kibana (ELK Stack) to visualize the proposed detection system.  

\subsubsection{Clustering Using a Similarity Measure Approach Based on Semantic Analysis of Adversary Behaviors, 2020,~\cite{wang2020clustering}} This work describes a 3-layers approach to analyze the statistical features of APT groups. Those 3 layers are Goal, Behavior and Capability. Authors mainly concentrate on the statistical features of APT groups, and utilize this model to analyze optimized methods of using threat intelligence information. Next, they present a similarity measuring method to capture similarity degree considering different semantic relationships between groups. Finally, they considered the Girvan–Newman clustering algorithm~\cite{kernighan1970efficient} to find community groups. In their experiments, they build a knowledge network for the behavior analysis leveraging the \mitre\ \attack\ data, and finally, they considered the similarity among the groups by building a knowledge network from the collected data.

\subsubsection{\mitre\ ATT\&CK-driven Cyber Risk Assessment, 2022,~\cite{ahmed2022mitre}} Authors propose a hybrid cyber risk assessment methodology based on (i) vulnerability-oriented approach, (ii) asset/impact-oriented approach, (iii) threat-oriented approach. This work combines the \attack\ repository with attack graphs to calculate attacks' occurrence probability and their potential success rate. For this purpose, first, it identifies business-critical objectives and corresponding threats. Then, it identifies special threat and environmental parameters to analyze the threat and its potential impacts. To summarize, the proposed methodology has 3 main steps: (i) organization modeling, (ii) threat modeling, and finally, (iii) impact assessment. The evaluation is supported by a case study of a health center located in the United Kingdom to conduct a series of risk analysis tasks against two adversary groups, i.e., Lazarus and menuPass.

\subsubsection{HOLMES: Real-time APT Detection through Correlation of Suspicious Information Flows, 2019,~\cite{milajerdi2019holmes}} Authors propose an APT detection system, namely HOLMES, that correlates tactics, techniques, and procedures deployed to carry out by each APT. This work proposes an APT detection system (HOLMES) that extracts suspicious information during cyberattacks. HOLMES starts with host logs or audit data, such as Linux auditd or Windows ETW data, and by leveraging the \attack\ matrix, it issues a detection alert mapping the phases of an ongoing APT campaign. Then, it generates a high-level scenario graph (HSG) that represents the attacker’s malicious actions in real-time. HSG is employed by security analysts for designing real-time cyber-response operations. The authors have evaluated HOLMES on a dataset provided by DARPA Transparent Computing program that employed a professional red-team simulating nine real-world APT threats on a
dynamic network consisting of different platforms. Based on the obtained results, they believe that HOLMES is able to detect and analyze APT campaigns with an acceptable accuracy and low false alarm rate. Finally, HOLMES can be deployed to serve as a real-time intrusion detection tool. 

\subsubsection{A machine learning framework for investigating data breaches based on semantic analysis of adversary’s attack patterns in threat intelligence repositories, 2019,~\cite{noor2019machine}} This work proposes a machine learning based framework that detects suspicious and malicious activities based on its previously observed attack patterns used as its training dataset. The proposed solution semantically identifies and correlates threats and TTPs provided by popular knowledge-bases, e.g., \mitre\ ATT\&CK, with corresponding detection mechanisms to build a semantic network. 
Such a dynamic semantic network is employed by security experts to detect various attacks by forming probabilistic relationships among threats and TTPs. For training the proposed framework, the authors use a TTP taxonomy dataset, and they use threat artifacts reported in threat reports for its performance evaluation. Finally, the authors have conducted experimental tests based on a financial malware case study. Their results show 92\% accuracy rate and significantly low false positive rate.  

\subsubsection{Optimal Machine Learning Algorithms for Cyber Threat Detection, 2018,~\cite{farooq2018optimal}} This work presents multiple examples about leveraging machine learning analytic to enhance cybersecurity monitoring and analysis. The authors propose machine learning techniques as an ideal choice to design and develop context-aware detection mechanisms helping to reduce false positive rate. To this aim, they firstly analyze multiple taxonomies to choose a standard cyberthreat knowledge-base. They select \attack\ and CAPEC to design their methodology. Next, they analyze applications of supervised, semi-supervised and unsupervised techniques in threat detection, and they conclude that semi-supervised algorithms, e.g., One-Class SVM, are significantly efficient and cost-effective enabling Security Operation Centers (SOC) analysts conduct new detection methods and discovering new indicators of compromise (IoCs).  

\subsubsection{Artificial Intelligence empowered Cyber Threat Detection and Protection for Power Utilities, 2019,~\cite{hasan2019artificial}} This work analyse how Artificial Intelligence can enhance the performance and efficiency of traditional cyberattacks detection systems. More specifically, the authors investigate how Machine Learning (ML) techniques can detect different attack stages of APT. \mitre\ \attack\ framework is employed for this purpose. They believe that AI-enabled detection systems can detect anomalies in wider patterns of adversary activities and present more comprehensive and dynamic detection capabilities. IBM Watson\footnote{https://www.ibm.com/watson} is adopted for cyber analysis tasks and detecting security threats. Finally, authors consider as a main use case the power utilities SCADA systems. 

\subsubsection{Automating Cyber Threat Hunting Using NLP, Automated Query Generation, and Genetic Perturbation, 2021,~\cite{karuna2021automating}} Authors propose AI-based techniques for the automation of DSL entries in support of threat detection. This work introduces the WILEE system which is an automated attack hunting system translating high-level abstract cyberthreat reports into multiple detailed concepts. Custom domain specific language (DSL) is used to represent both high-level and corresponding detailed concepts. WILEE uses such concepts to automatically generate queries to validate the hypotheses tied to the potential adversarial activities. Next, two AI components of WILEE that automate its query generation for hunting are described in details: (i) a hunt component, called Malmo, that automates extraction and translation of known threat descriptions from \mitre\ \attack\ into machine digestible format using NLP techniques, (ii) a genetic programming hunt component, called Genetic Perturbations Engine (GPE), as a generator of implementations of threats.

\subsubsection{GroupTracer: Automatic Attacker TTP Profile Extraction and Group Cluster in Internet of Things, 2020,~\cite{wu2020grouptracer}} Authors propose a framework, called GroupTracer, for automatically observing and predicting complex IoT attacks based on \attack\ matrix. The proposed framework identifies attacks activities by leveraging IoT honeypots and it retrieves complementary information from corresponding logs. Next, it maps the attack behaviors to the \attack\ matrix to perform automatic TTP profiles extraction. Finally, using hierarchical clustering techniques and based on four feature groups (TTP profiles, Time, IP, URL) and 18 characteristics derived from log entries, it clusters attack behaviors and generates attack trees for each cluster in order to discover potential groups behind complex attacks.  In the experiment phase, they compare their proposed algorithm with two baseline algorithms by evaluating 395,006 log entries from 3,025 IPs. Based on the achieved results, they claim that GroupTracer can reach to highly acceptable performance as the Calinski–Harabasz index reaches 3416.93.

\subsubsection{DMAPT: Study of Data Mining and Machine Learning Techniques in Advanced Persistent Threat Attribution and Detection, 2021,~\cite{charan2021dmapt}} This work analyses machine learning and data mining techniques performance and efficiency for APT malware detection and profiling. First, it compares different APT detection techniques and highlights their research gaps to be addressed by cybersecurity community. The authors believe that even though researchers have proposed different APT detection techniques, there are significant shortcomings to be addressed in order to enhance current detection mechanisms. The identified challenges are: (i) lack of qualified benchmark datasets for training and evaluation, (ii) continuous changes in TTP usage by APTs resulting high false positive rate. Consequently, qualified and comprehensive benchmark datasets and well-designed robust machine learning models are needed in order to characterize and detect APTs.

\subsubsection{Open Source Intelligence for Malicious Behavior Discovery and Interpretation, 2021,~\cite{huang2021open}} Authors propose a \mitre\ \attack\ based malicious behavior analysis system (MAMBA) incorporating \attack\ knowledge into neural network models for MS Windows malware detection. This work exploits Open Source Intelligence (OSINT) for the analysis of several malware. The goal is to discover malicious activities and corresponding TTPs by analyzing the execution trace associated with the malware under the Windows operating system. For this purpose, they follow 3 main phases: (i) OSINT for cyber threat intelligence, (ii) malicious behavior discovery, and finally, (iii) malicious behavior explanation. Next, they propose the MAMBA system addressing the aforementioned aspects. MAMBA uses a well-designed neural network model to extract TTPs from the \attack\ matrix and correlate them with corresponding malware API execution call sequences. For regular synchronization with \attack\ updates, MAMBA automatically collects knowledge in an incremental process. Consequently, they authors believe that MAMBA is explainable, comprehensive and extendable with a significantly high detection rate compared to the other analyzed methods and approaches. To prove this, MAMBA has been evaluated based on four scenarios: (i) performance evaluation and comparison with other methods, (ii) ablation test, (iii) APT29 case study, and finally, (iv) resource and API locating. 

\subsubsection{Game-Theoretic Decision Support for Cyber Forensic Investigations, 2021,~\cite{nisioti2021game}} This work models the interaction between a cybersecurity analyst and an advanced attacker using a game-theoretic framework (Bayesian game). A graph of actions representing the investigator and the attacker moving across different hosts is generated. The graph edges represent TTPs among different network nodes and the graph nodes represent parties' decisions. Hence, using this modeling the investigator can identify the optimal policies by inspecting the cost and impact of every action.
Accordingly, a game-theoretic decision support framework is proposed trying to improve the performance of an investigator against a highly capable attacker. The authors believe that this is the first step towards using game-theoretic concepts for forensic investigations. As for the evaluation, they conduct a series of experiments by constructing a real-life case study based on: (i) threat reports collected from the \mitre\ \attack\ STIX repository, (ii) Common Vulnerability Scoring System (CVSS), and finally, (iii) consultations with cyber-security experts. They compare the achieved results with two other investigative methods (Uniform and CSS) and three different types of attackers.

\subsubsection{Fingerprint for Mobile-Sensor APT Detection Framework (FORMAP) Based on Tactics Techniques and Procedures (TTP) and MITRE, 2022,~\cite{al2022fingerprint}} This work proposes a fingerprint for mobile-sensor APT detection framework (FORMAP) based on the correlation between the \mitre\ framework, the mobile sensors and the attack trees. The goal of the proposed framework, namely FORMAP,  is to improve the security awareness to detect APT attacks on smartphones. The main objectives of this work are: (i) studying smartphone sensors (inertial, positioning and ambient) vulnerabilities exposed to APT attacks, (ii) designing a set of attacks trees for Mobile APT based on \mitre\ \attack\ framework, and finally, (iii) proposing a smartphone APT detection framework. The authors believe that FORMAP can be a ground knowledge for researchers to better understand mobile phones TTPs. 

\subsubsection{Evolution of IoT Linux Malware: A \mitre\ \attack\ TTP Based Approach, 2021,~\cite{chierzi2021evolution}} 
Regarding the significant increase on the attacks targeting Internet of Things (IoT) devices, this research proposes an approach to keep track of the evolution path of techniques and capabilities employed by IoT linux malware. For this purpose, they leverage \mitre\ \attack\ matrix to discover and characterise current threats (also covering pre- and post-exploitation aspects) and provide useful insights. In their analysis environment, they analyze 14 unique IoT Linux malware families collected recently. Based on the achieved results, they claim that ransmoware are more static respect to botnet malware, which indeed changes quickly and can be employed by different kinds of adversary techniques.

\subsubsection{Identifying \attack\ Tactics in Android Malware Control Flow Graph Through Graph Representation Learning and Interpretability, 2021,~\cite{fairbanks2021identifying}}. This research proposes an approach to automate identifying and locating TTPs in a Control Flow Graph (CFG) by using Graph Machine Learning techniques on Android Malware. The authors believe that identifying sub-graphs in the malware CFG provides insights about malware behavior and corresponding mitigation strategies. For this purpose, from a CFG, the sub-graph representing malware executable execution flow is selected (using attribution techniques) and associated with corresponding TTPs. They use Graph Neural Networks and SIR-GN node representation learning technique to analyze CFGs and generate a model that classifies associated TTPs. Based on their experiments, they believe that the proposed methodology outperforms graph isomorphic networks and graph attention networks.
\\\\
{\bf Wrap up on {\em Behavioural analytic.}} This class of use cases focuses on providing the most accurate model of the adversary in terms of sequence of malicious actions that can be done against a target. The \attack\ matrix can be very helpful in mainly two ways: (i) as a source of information and (ii) as a support to behavioural modelling. The former considers the \attack\ matrix as a database of techniques---as indeed it is---while considering hidden relations and extracting correlations to perform a better adversarial modelling to improve the identification of the APT behind the attack. The latter considers the \attack\ matrix as a support tool for different purposes, e.g., matching the adversarial behaviour extracted from real measurements, abstracting from a pattern of actions in a specific scenario to a more general behaviour, generalization of results, or finally, driving the analysis of real data.

%\begin{figure}
%    \centering
%    \includegraphics[width=0.7\columnwidth]{figures/behavioral_analytics.pdf}
%    \caption{Behavioral analytic with the \attack\ matrix: We identified three main relations with the \attack\ matrix, i.e., Input, Support, and Comparison. The \attack\ matrix can be used as (i) a data input---being a database of techniques, (ii) a support tool for driving the description of the adversarial behaviour, or finally, (iii) as a comparison criteria---being itself a tool for adversarial behaviour categorization.} 
%    \label{fig:behavioral_analytics}
%\end{figure}

\begin{figure}
    \centering
    \includegraphics[width=0.9\columnwidth]{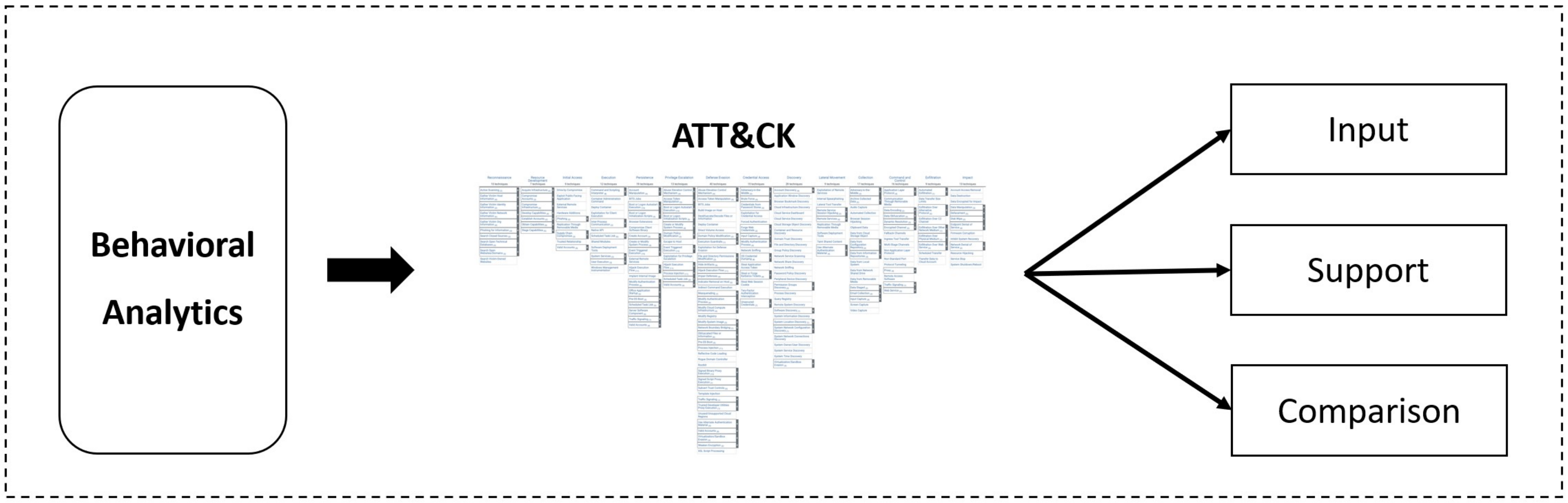}
    \caption{Behavioral analytic with the \attack\ matrix: We identified three main relations with the \attack\ matrix, i.e., Input, Support, and Comparison. The \attack\ matrix can be used as (i) a data input---being a database of techniques, (ii) a support tool for driving the description of the adversarial behaviour, or finally, (iii) as a comparison criteria---being itself a tool for adversarial behaviour categorization.} 
    \label{fig:behavioral_analytics}
\end{figure}

Finally, we found one only paper performing a comparison between the \attack\ framework and STRIDE, and we decided to report it as standalone. Although the proposed categorization for the behavioral analytic might be blurry for some of the papers, we try to extract the main relation in order to provide a more clear analysis when reporting the paper as it follows:

\begin{itemize}
    \item[-] {\em Input:} \cite{rawan2020}, \cite{wang2020clustering}, \cite{ahmed2022mitre}, \cite{milajerdi2019holmes}, \cite{farooq2018optimal}, \cite{nisioti2021game}, \cite{chierzi2021evolution}, and \cite{fairbanks2021identifying}.
    \item[-] {\em Support:} \cite{munaiah2019characterizing}, \cite{jadidi2021threat}, \cite{toker2021mitre}, \cite{noor2019machine}, \cite{hasan2019artificial}, \cite{karuna2021automating}, \cite{wu2020grouptracer}, \cite{charan2021dmapt}, \cite{huang2021open}, and \cite{kadhimi2022}.
    \item[-] {\em Comparison:} \cite{straub2020modeling}.
\end{itemize}

%Figure~\ref{fig:behavior_pie} shows the identified categorization, i.e., Input, Comparison and Support, for the behaviour analytic use case.

% \begin{figure}
%     \centering
%     \includegraphics[width=\columnwidth]{figures/behavior_pie.pdf}
%     \caption{Behaviour analytic categorization. Input: the \attack\ matrix is used as input database; Support: the \attack\ matrix supports the behavioural modelling of the adversary, and finally, Comparison, the \attack\ framework is used as a reference model for comparison. 
%     \label{fig:behavior_pie}
% \end{figure}

\subsection{Red teaming}

{\em Red teaming} represents the set of practices of challenging a target system in order to test its security and privacy robustness. In this section, we consider the red teaming use case and we study how the current state of the art leverages the \attack\ framework to enhance such activity.

\subsubsection{Threat Modeling Based on the \mitre\ Enterprise \attack\ Matrix, 2022,~\cite{xiong2022cyber}} Authors uses the \mitre\ Enterprise \attack\ matrix to propose a threat modeling language, called {\em enterpriseLang}, for enterprise security. The proposed threat modeling language which is a Domain Specific Language (DSL), is designed based on the Meta Attack Language (MAL) framework~\cite{johnson2018meta} and concentrates on representing system assets, asset associations, attack phases, and defense strategies. The attack phases in {\em enterpriseLang} represent \mitre\ \attack\ matrix adversary techniques. For modeling enterprise IT systems, an entity-relationship model is embedded in the language enabling attack simulations on the system model instances. 
From design perspective, {\em enterpriseLang} has six steps: (i) problem definition \& Motivation, (ii) Defining general and specific Objectives, (iii) Design \& Development, (iv) Demonstration, (v) Evaluation, (vi) Communication. For evaluation, the proposed language is tested based on a group of unit and integration tests. More specifically, {\em enterpriseLang} is used to model two known real-world attack scenarios: the Ukraine cyber attack and the Cayman National Bank cyber heist.

\subsubsection{SoK: A Survey of Open-Source Threat Emulators, 2022,~\cite{zilberman2020sok}} Authors provides a survey reviewing and comparing 11 open-source threat emulators in order to verify whether \mitre\ \attack\ tactics and techniques can be performed and tested by these emulators. In this survey, the authors propose a methodology to quantitatively and qualitatively compare threat emulators with respect to a list of criteria and present a taxonomy that illustrates threat emulators quality. The main objective of this work is providing guidelines for security teams to choose the most appropriate cyberthreat emulator depending on the target context and application. 
The investigated threat emulators are: PowerSploit, APTSimulator, Uber’s Metta, CALDERA, Infection Monkey, DumpsterFire, Atomic Red Team, Metasploit, Red Team Automation, BT3 and Invoke-Adversary. Their obtained results demonstrate that these four open-source threat emulators are outperforming others: RTA, Atomic Red Team, Metasploit and CALDERA.

\subsubsection{Probabilistic Attack Sequence Generation and Execution Based on \mitre\ \attack\ for ICS Datasets, 2021,~\cite{choi2021probabilistic}} Authors propose an automatic attack sequence generation approach consistent with \mitre\ \attack\ tactics and techniques in order to define and deploy an attack dataset. In this research, the authors consider three aspects for a qualified attack sequence generation: (i) reproducibility, (ii) diversity, (iii) reality. Accordingly, they propose an attack sequence generation approach that uses Hidden Markov Models (HMM) as its core component. The main characteristics of this approach are: (i) representable attack sequence, (ii) probabilistic attack sequence, and finally, (iii) practical attack sequence. The authors have evaluated their proposed approach based on 3 different case studies covering general and specific ICS infrastructures. 

\subsubsection{powerLang: a probabilistic attack simulation language for the power domain, 2020,~\cite{hacks2020powerlang}} Authors propose {\em powerLang} as a Meta Attack Language (MAL) based Domain-Sepcific-Language (DSL) that can model both power domain information technology (IT) and operational technology (OT) infrastructures. {\em powerLang} enables power domain practitioners to model their IT/OT architectures and accordingly simulate the activities of malicious entities. This helps them to detect possible security breaches and improve their infrastructure's security. Authors reuse two exiting MAL-based DSL: (i) coreLang~\cite{katsikeas2020attack}, for modeling common information technology infrastructures, and (ii) sclLang\footnote{https://github.com/mal-lang/SCL-Lang}, for modeling the internal schema of substations. Next, in order to cover the inconsistencies between the information technology infrastructures and other technical infrastructures, they introduce icsLang that serves as a tool for modeling OT environments. In their experiment, they evaluate the proposed language against a known cybersecurity attack, simulating the Ukrainian electric power grid attack scenario. 

\subsubsection{Attack Specification Language: Domain Specific Language for Dynamic Training in Cyber Range, 2021,~\cite{arshad2021attack}} This research leverages the \mitre\ \attack\ framework to propose an Attack Specific Language (ASL) that provides a unified representation for all the cyber-threat scenarios. ASL provides information and knowledge about attack techniques in a brief format that streamlines and automates the malicious activities of threat and challenge execution. It facilitates designing superior and dynamic threat scenarios by providing comprehensive specification methods. Its main features are of easy implementation, seamless integration and process improvement. Authors use the game-theory modeling to provide a formal proof for an improved ASL dynamic training and customization.

\subsubsection{Analysis Of Automated Adversary Emulation Techniques, 2017,~\cite{applebaum2017analysis}} This research propose a test-bed for simulating attackers behavior and activities in Windows-based enterprise networks. After conducting a series of experiments, authors prove that an automated attacker outperforms non-automated ones. Their main objective is expanding the automation to the tactical level despite the existing difficulties and challenges. The major components of the proposed simulation environment are: objects or entities, entity-relationships, and control settings. Each simulation can be considered as a game having 3 distinct players: an agent simulating user activities, an attacker and a defender. Overall, sixteen scenarios were created including variations of agent actions, vulnerabilities, connections and detection frequencies. They achieved promising initial results in their experiments automating the entire emulation process. 

\subsubsection{A review of threat modelling approaches for APT-style attacks, 2021,~\cite{tatam2021review}} This work proposes a literature review investigating existing threat modeling approaches (49 current studies) and building a threat modeling knowledge-base. The main objectives of this research is to define and discover threat modeling challenges, strengths, potential gaps and corresponding enhancements aiming to improve threat modeling efficiency when modeling more complicated attacks, such as the ones coming from Advanced Persistent Threats (APTs). The reviewed research works and approaches are related to different levels of an organization but mainly related to system and data-centric software design, development and operations. Authors claim that there is no particular threat modeling solution covering various assets, threats, vulnerabilities, systems and data. There are a number of features helping to determine which approach to be taken: (i) identify agents, their motivations and capabilities, (ii) correlate threat intelligence information from multiple sources, (iii) identify critical assets and controls, (iv) include stakeholders from different levels, (v) determine system-wide environmental boundaries, (vi) define authentication and authorisation policies, and finally, (vii) select a comprehensive approach that satisfies pre-defined requirements and consists with the underlying audience and environment. 

\subsubsection{Towards Dynamic Threat Modelling in 5G Core Networks Based on \mitre\ \attack\, 2021,~\cite{pell2021towards}} Authors propose an approach to extend \mitre\ \attack\ framework by adding a list of 5G related cyber-threat techniques. This research investigates the knowledge sharing between early 5G networks threat analysis and adversarial TTPs of \mitre\ \attack\ framework, and it proposes solutions to bridge these gaps. It analyzes the knowledge and theoretical shortcomings of current 5G technology enablers (e.g., Software-defined networking (SDN) and network functions virtualization (NFV)). The main objectives are: (i) identifying 5G Core Networks (5GCN) threat modeling and risk assessment, (ii) adding 5GCN related adversarial techniques into the \mitre\ \attack\ framework. Identifying 5GCN adversarial techniques and mapping them to the \mitre\ \attack\ framework tactical stages can empower this framework to serve future 5GCN cyber-threat and cyber risk analysis tasks. 

\subsubsection{From TTP to IoC: Advanced Persistent Graphs for Threat Hunting, 2021,~\cite{berady2021ttp}} A formal method modeling cybersecurity concepts from both attacker and defender point of view is proposed in~\cite{berady2021ttp}. For this purpose, two persistent graphs are generated: (i) the execution of attacker's procedures mapped to the \mitre\ \attack\ framework, (ii) the exposed resources during the execution of attacker's procedures. The graphs help to reduce false positive alarm rates for defenders. Additionally, the proposed approach enhances the defender's knowledge-base by adding new IoCs enabling proactive threat detection. The authors believe that the mutual inference for both attacker and defender in order to improve their knowledge is necessary to reach their goals (in particular for the defender). The proposed approach has been evaluated using an attack campaign mimicking APT29 scenario. 
\\\\
{\bf Wrap up on {\em Red teaming.}} The red teaming use case comprises all the papers that leverage the \attack\ framework to develop, deploy or implement red teaming activities, i.e., malicious activities aiming at testing the security and privacy robustness of a target infrastructure. We identified 3 sub-categories, i.e., {\em real attacks}, {\em emulated attacks}, and finally, {\em threat (scenario) modelling}. Real attacks (\cite{choi2021probabilistic}) involve papers dealing with real attack patterns while considering the \attack\ matrix to design the attack itself. Emulated attacks (\cite{zilberman2020sok} \cite{hacks2020powerlang}, \cite{applebaum2017analysis}) involve the study, design and development of tools for the delivery of emulated attacks while the \attack\ framework is generally considered as the source for the best-known tactics and techniques for performing the attack. Finally, threat modelling (\cite{xiong2022cyber}, \cite{arshad2021attack}, \cite{tatam2021review}, \cite{pell2021towards}) involves the development of languages or frameworks for the in-depth description and analysis of the threats.

\begin{figure}
    \centering
    \includegraphics[width=0.9\columnwidth]{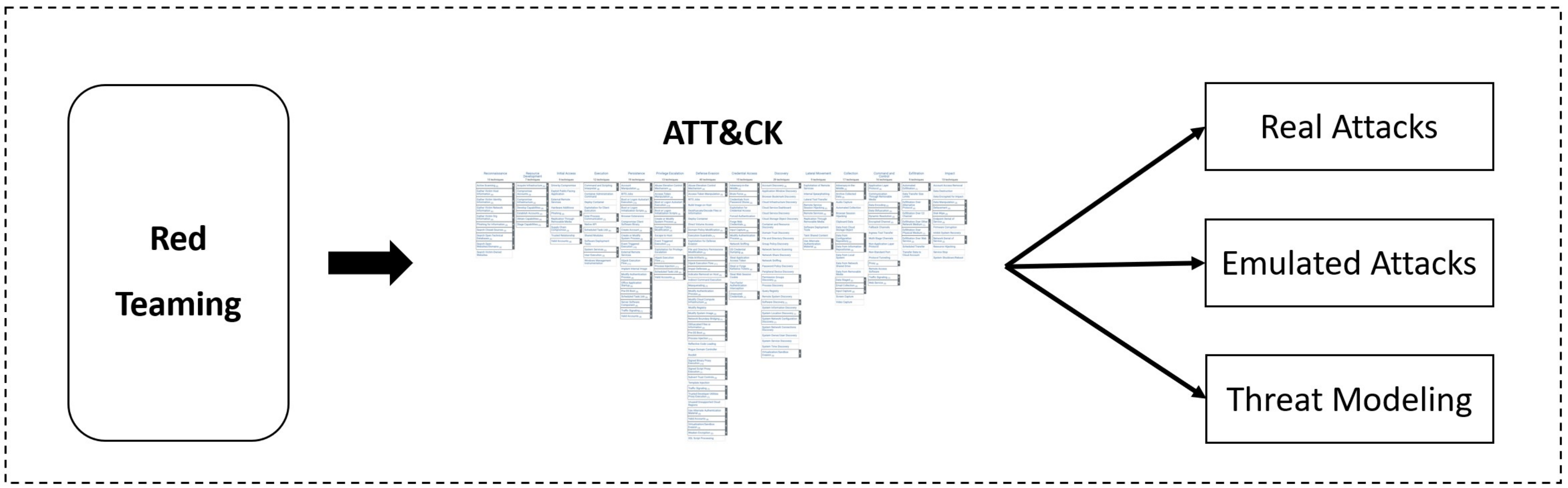}
    \caption{Red teaming with the \attack\ matrix: We identified three main sub-categories where the \attack\ matrix is used for supporting red teaming activities. Indeed, the tactics and techniques can be leveraged to generate real attacks or environments with emulated ones. Threat modelling can also be supported by the \attack\ matrix for the definition of new languages and framework that better describe threats.} 
    \label{fig:red_teaming}
\end{figure}

\subsection{CTI Enrichment}
The \attack\ matrix is a database of techniques previously adopted by APTs to launch attacks. When such techniques become linked together to create an attack chain and they can be correlated with other ones to identify and characterize APTs, they become a core component of the Cyber Threat Intelligence (CTI)---the \attack\ framework can be considered as a ground-truth for sharing information among organizations, thus enriching the internal intelligence about the current threat landscape.

\subsubsection{Assessing \mitre\ \attack\ Risk Using a Cyber-Security Culture Framework, 2021,~\cite{georgiadou2021assessing}} Authors propose to correlate a well-defined set of features including both organizational and individual features with the \mitre\ \attack\ framework TTPs and vulnerability databases. In this work, a multidisciplinary security culture framework covering attack patterns of the \mitre\ \attack\ matrix, designed for critical infrastructures is being evaluated. This framework can be applied to a wide range of applications, such as organizing security protocols and improving security evaluation results, etc. The main objective of the authors is to link a holistic set of cyber-security culture factors with the tactics and threats listed in \mitre\ \attack. The proposed cyber-security culture framework evaluates organizations security culture from different perspectives including human and business related parameters. For evaluation, they analyze the framework based on two application scenarios (simple and complex) to demonstrate its adaptability with various type of organizations having different sizes and structures. 

\subsubsection{Cyber Threat Dictionary Using \mitre\ \attack\ Matrix and NIST Cybersecurity Framework Mappting, 2020,~\cite{kwon2020cyber}} Authors propose an attack-defense mapped framework, called Cyber Threat Dictionary (CTD), linking \mitre\ \attack\ matrix to NIST framework. Authors believe that CTD can be used re-actively or proactively within critical infrastructures in order to: (i) identify security gaps to link them to \attack\ techniques, (ii) develop suitable mitigation and detection techniques, (iii) correlate critical cyber-attack techniques and the Facility Cybersecurity Framework (FCF)\footnote{https://facilitycyber.labworks.org/} controls, (iv) vulnerability prioritization based on the FCF and the correlation matrix, (v) develop appropriate prevention solutions. The two major components of CTD are its search engine which provides details and solution to deal with an attack, and it solution suggestion component which issues appropriate solutions. 

\subsubsection{Automated Retrieval of \attack\ Tactics and Techniques for Cyber Threat Reports, 2020,~\cite{legoy2020automated}} This research proposes an automated cyber threat report analysis tool, called rcATT to extract valuable security information from Cyber Threat Reports (CTRs). In this research work, in order to automate information extraction from CTRs, the authors investigate mutli-label text classification techniques and map them to the \mitre\ \attack\ tactics and techniques.  Accordingly, they present rcATT, a solution for predicting \attack\ tactics and techniques from input CTRs and providing the results in STIX format. rcATT provides the possibility for users to interactively adjust or correct the classification results and improve its performance over time. For evaluation, they test and compare rcATT's features with available open-source solutions and explain its acceptable performance and efficiency.    

\subsubsection{Learning APT Chains from Cyber Threat Intelligence, 2019,~\cite{husari2019learning}} Authors propose a Natural Language Processing (NLP) based approach to characterise an APT attack actions temporal relationship , and map them to \mitre\ \attack\ and STIX-2 frameworks. This research work presents its preliminary results of capturing temporal relationship between threat Indicators of Compromises (IoCs) provided by cyber threat intelligence. For this purpose, first, they design a set of Regular Expressions (regex) to extract IoCs from text data. Next, they develop an NLP typed dependency rule-set to list the temporal relationship between the IoCs in cyber-threat reports and map them to the five stages of kill chain (Deliver, Exploit, Command\&Control, Stage, Exfiltrate) associated with the threat IoCs. In the experimentation phase, they process 5000 threat reports from Symantec and Fireeye CTI blogs\footnote{https://www.mandiant.com/cyber-threat-intelligence-visualization} and analyze them to generate STIX machine-readable constructs. These constructs can be processed by firewalls and Intrusion Detection Systems (IDS). According to the results, they belive that the IoCs extracted from their dataset represent high temporal order correlation helping security solutions to detect attackers activities based on a sequence of temporal events and corresponding IoCs. This helps to improve detection accuracy by reducing the false positive alarm rate generated by IDSs. 

\subsubsection{Linking CVE’s to \mitre\ \attack\ Techniques, 2021,~\cite{kuppa2021linking}} This research proposes a neural-network based model to automate maping CVE’s to \attack\ techniques. The authors to address the lack of labeled data, propose an unsupervised labeling approach that extracts and analyzes phrases from security reports and \attack\ technique descriptions. Accordingly, they were able to map 17 techniques and create a knowledge-base including 150 attack scenarios and 50 mitigation procedures. The knowledge-base includes both adversary and defender views for any input CVE. Next, using the proposed Multi Label Text Classifier (MLTC), they correlate CVE's with \attack\ techniques. For evaluation, they use a dataset covering past 10 years CVEs and compare the proposed approach performance with models, such as Term Frequency-Inverse Document Frequency (TFIDF) based SVM multi-label classifier, Bi-direction Long Short-Term Memory (BI-LSTM) and Attention-based BI-LSTM.

\subsubsection{Expansion of ICS Testbed for Security Validation based on \mitre\ \attack\ Techniques, 2020,~\cite{choi2020expansion}} Authors propose an extended testbed to collect and monitor events during a ICS attack, and map them to the \mitre\ \attack\ matrix. In this research, the authors main objective is to collect and monitor information during an ICS attack and map them to \mitre\ \attack\ framework. Moreover, they propose a framework to simulate various sophisticated cybersecurity attack scenarios. For this purpose, they extend an existing testbed called Hardware-In-the-Loop (HIL) based Augmented ICS (HAI). The authors mainly emphasize on data diversity, data ingestion, attack scenario and time synchronization. 

\subsubsection{Linking Common Vulnerabilities and Exposures to the \mitre\ \attack\ Framework: A Self-Distillation Approach, 2021,~\cite{ampel2021linking}} Authors propose a self-distillation model called CVE Transformer (CVET) aiming to label CVEs based on information coming from the \mitre\ \attack framework. CVET includes a fine-tuned distillation model that is the result of using a pre-trained language model called RoBERTa. The proposed methodology has 3 main components: (i) data collection \& pre-processing, (ii) the CVE Transformer (CVET) Architecutre, and (iii) benchmark experiments. The proposed model has been evaluated based on a gold-standard dataset and its performance has been compared with classical machine learning (e.g., Random Forest and SVM), deep learning (e.g., RNN and Transformer) and pre-trained NLP models, such as BERT and RoBERTa. As future direction, the authors intend to connect and correlate their framework with the CAPEC list and the NIST framework. 

\subsubsection{Cyber Threat Intelligence Model: An Evaluation of Taxonomies, Sharing Standards, and Ontologies
within Cyber Threat Intelligence, 2017,~\cite{mavroeidis2017cyber}} This work proposes a CTI model enabling security analysts to analyze cyber-threats. In this research, first the authors discuss about a significant need for an ontology for CTI, and list existing difficulties toward proposing such an ontology, such as lack of standard data representation, comprehensive terminology list and their layered hierarchical relationships. Next, they analyze and evaluates existing taxonomies, standards and ontologies. Finally, they propose a CTI model which tries to improve expressiveness of existing taxonomies and ontologies. For evaluation, they use three criteria: (i) covered data and concepts, (ii) correlation with other taxonomies and ontologies, (iii) how easily can be analyzed and investigated. Their results show that proposed ontologies must follow Ontology Web Language (OWL) constraints and restrictions that guarantee high reasoning abilities.

\subsubsection{TTPDrill: Automatic and Accurate Extraction of Threat Actions from Unstructured Text of CTI Sources, 2017,~\cite{husari2017ttpdrill}} Authors propose an automated and context-aware CTI analytic approach to extract adversarial patterns (TTPs) from accessible CTI repositories and issue appropriate responses. Authors utilizes CAPEC and \attack\ knowledge-bases to propose a comprehensive attack-response ontology to be used for analyzing cyber-threats and corresponding responses. In addition, they develop a novel Natural Language Processing (NLP) based approach to search and capture structures TTPs. Finally, they connect these components to build a complete threat actions analysis tool, called TTPDrill, mapping each action to an appropriate TTP. TTPDrill is able to analyze structured and unstructured CTI reports and extract malicious activities and corresponding TTPs in an standard format, such as STIX 2.1. For evaluation, they use a dataset of 17000 cyber-threat reports provided by Symantec. The evaluation results demonstrate that TTPDrill's detection precision and recall goes over 82\%.

\subsubsection{Integrating Cyber-D\&D into Adversary Modeling for Active Cyber Defense, 2016,~\cite{stech2016integrating}} This research introduces cyber Denial and Deception (cyber-D\&D) that uses an experimental and engineering paradigm to combines security research and experiments into cyber-D\&D actions. In other words, it discusses how to integrate cyber-D\&D tools and TTPs into threat modeling knowledge-bases, such as \mitre\ \attack\, aiming to provide active cyber defense (ACD) recommendations for critical environments and infrastructures. For this purpose, they describe the key elements required for this integration: (i) cyber threat intelligence, (ii) systems and network sensors, (iii) intrusion detection and analysis, (iv) defender mitigation, (v) Red-Blue experimentation. They believe that a defensive cyber-D\&D integrates standards, threat repositories, operational resources and active defense systems. 

\subsubsection{Towards a Definition of Cyberspace Tactics, Techniques and Procedures, 2017,~\cite{maymi2017towards}} This paper proposes a set of models for offensive cyberspace operations (OCO) allowing autonomous detection of cyber-attacks. Authors provide the definitions for cyberspace operations supported by mathematical modeling to improve their efficiency and usefulness for advanced data analytics applications. Finally, these definitions are proved to be consistent to model a sequence of real-world cyberspace activities conducted by APT28\footnote{https://attack.mitre.org/groups/G0007/}. 

\subsubsection{On the use of Cyber Threat Intelligence (CTI) in Support of Developing the Commander’s Understanding of the Adversary, 2019,~\cite{parmar2019use}} This research describes the NATO Allied Command Transformation (ACT) and the NATO Communication and Information Agency (NCI Agency) experiments that concentrate on using deception techniques, such as honeypots to capture and analyze attackers activities data. In particular, the \mitre\ \attack\ framework has been utilized during the experiments to identify known adversary patterns and to be correlated with the collected traffic in order to provide required information for responsible Commanders. The overall goal was achieved by conducting three test cases: (i) validating the importance of honey pots data, (ii) validating the relevancy of captured data with \mitre\ \attack\ framework, and finally, (iii) validating the operational context based on the results of previous steps. 

\subsubsection{Attack Hypothesis Generation, 2019,~\cite{elitzur2019attack}} Authors propose an Attack Hypothesis Generator (AHG) to generate hypothesis regarding cyber-attacks existing in the underlying context. AHG leverages threat intelligence knowledge-bases (e.g., \mitre\ \attack) to issue useful complementary information about the equipment and strategies used by attackers during a cyber-attack. For this purpose, it complements threat intelligence graph by adding information coming from recommendation provision techniques to identify ongoing attacks features and attributes resulting in actionable investigative items. More specifically, authors use machine learning techniques (both supervised and unsupervised) for prediction and collaboration tasks. In summary, AHG's contributions are: (i) utilizing CTIs to enhance attack hypotheses, (ii) enhancing hypotheses accuracy, (iii) suggesting actionable items. One of the major outcomes and benefits of AHG is improving automated forensic analysis tasks and procedures. For AHG performance evaluation, the authors prepared a hypothesis simulation environment to analyze and compare experts' hypothesis with AHG ones. The results demonstrate significant accuracy increase on the experts' hypothesis that can lead toward high-end and sophisticated investigations and inspections. 

\subsubsection{Automatically Attributing Mobile Threat Actors by Vectorized \attack\ Matrix and Paired Indicator, 2021,~\cite{kim2021automatically}} Authors propose an automated methodology to classify mobile malware using \mitre\ \attack\ framework. For this purpose, a mathematical modeling of the \attack\ matrix is proposed to analyze and compare Indicators of Compromise (IoC) and provide complementary insights. The main objective is to automate the task of classifying mobile threat related TTPs while decreasing the false positive rate. This can lead to process a significantly high volume of Mobile incidents TTPs that is very useful for the field. In their experiments, they use the proposed approach to evaluate 12 threat actors and 12 malware. Based on their results, the authors believe that this approach can appropriately handle large-volume cyber-threat analysis tasks that represents its potentials to be used in various threat intelligence organizations.

\subsubsection{Data-Driven Decision Support for Optimizing Cyber Forensic Investigations, 2021,~\cite{nisioti2021data}} This paper introduces DISCLOSE: a decision support platform for performing advanced forensic analysis tasks. It uses popular knowledge-base, such as \mitre\ \attack\ TTPs, as threat intelligence information to characterise threats and their behaviors/activities. Consequently, DISCLOSE empowers the security analysts to conduct efficient investigation of probabilistic relations among attacks actions. For evaluation, the authors verify DISCLOSE performance based on a case study that includes 9 cyber-threat scenarios (3 incidents and corresponding triggering actions) mapped to 31 TTPs. 

\subsubsection{Price TAG: Towards Semi-Automatically Discovery Tactics, Techniques and Procedures of E-Commerce Cyber Threat Intelligence, 2021,~\cite{wu2021price}} The main objective of this research is semi-automatically extracting TTPs from e-commerce threat intelligence corpora and mapping them with \mitre\ \attack\ framework. One main challenge is how to efficiently analyze and investigate a significantly high volume of e-commerce data to extract hidden CTIs. For this purpose, authors leverage NLP techniques (e.g., topic modeling and named entity recognition) to propose an approach called TTP Semi-Automatic Generator (TAG) performing e-commerce TTP extraction. In their experiments for TAG evaluation, they evaluate TAG using a dataset of 229,729 e-commerce security incidents. Their results show that TAG was able to extract 6,042 TTPs with 80\% precision proving that TAG has the capability to analyze highly complex and evolving threats. Moreover, they found a number of emerging 0-day attack patterns and corresponding CTIs showing exploited tools and strategies by attackers. They reported these findings to the Alibaba Group. As future work, they intend to address a limitation which is attackers can evade TAG by adjusting the operational model, such as using utility-preserving adversarial text. 

\subsubsection{TIM: threat context‑enhanced TTP intelligence mining on unstructured threat data, 2022,~\cite{you2022tim}} This research defines TTP intelligence representing all the TTP related information extracted from the under-analysis dataset. Hence, TTP classification can be represented as a phrase classification task. For TTP intelligence mining, first, they label a dataset of 10,761 cyber-threat reports that includes 6 TTP classes and 6,061 TTPs. Next, they propose a TTP Intelligence Mining (TIM) framework to extract TTP intelligence insights from a large-volume unstructured security dataset. TIM framework employs Threat Context Enhanced Networks (TCENet) to extract and classify TTP intelligence, and based on the authors claim, it is the first work performing the sentence level TTP classification. its results are generated in STIX 2.1 standard format to be shared and used with other tools. Their evaluation results show 0.941 accuracy which outperforms previous TTP classification approaches. Authors claim that adding TTP element features can even improve this classification accuracy. 

\subsubsection{Threat intelligence \attack\ extraction based on the attention transformer hierarchical recurrent neural network, 2022,~\cite{liu2022threat}} This work introduces Attention-based Transformer Hierarchical Recurrent Neural Network (ATHRNN) being an NLP attention-based model to automate extracting ana analyzing TTPs from unstructured CTI reports. ATHRNN uses \mitre\ \attack\ framework as its knowledge-base to extract TTPs. Authors present a Transformer Embedding Architecture (TEA) to retrieve semantic representations of CTIs and corresponding \attack\ metadata. Next, they present an Attention Recurrent Structure (ARS) to represent the  tactical and technical labels dependencies in \mitre\ \attack\ framework. Lastly, they use a joint Hierarchical Classification (HC) model to extract their target tactical and technical categories from cyber-threat datasets. 
To validate their proposed approach, authors build a benchmark dataset from threat reports generated from multiple sources. Their experiments results based on the gathered dataset, represent that they achieved 6.5\% and 8.2\% accuracy increase for Macro-F score and Micro-F score.

\subsubsection{Enhancing the STIX Representation of \mitre\ \attack\ for Group Filtering and Technique Prioritization, 2022,~\cite{zych2022enhancing}} This research enhances the \mitre\ \attack\ STIX representation to make it queryable based on various contextual information, such as activity groups motivations, the origin country, the target segments and locations, etc. For this purpose, authors discuss about the structure of \attack\ groups and details of the structures (e.g., MITRE’s STIX 2.1 representation). Next, they integrate the proposed enhancements to STIX 2.1 format of the \mitre\ \attack\. The enhancements mainly concentrate on the contextual information to enable constructing sophisticated queries and utilize the knowledge-base in a more detailed fashion. Hence, they employ STIX 2.1 objects to extract and represent the contextual information in a highly structured format. Finally, they demonstrate how to use the enhanced representation using programming (JSON compatible) and querying languages (e.g., MS SQL).

\subsubsection{Visualizing Interesting Patterns in Cyber Threat Intelligence Using Machine Learning Techniques, 2022,~\cite{ejaz2022visualizing}} This research proposes a machine learning based CTI reports visualization mechanism to illustrate hidden insights of threats and help analysts issue appropriate mitigation actions and recommendations. The main goal of the authors is to facilitate understanding the current cyber threat landscape and applying suitable mitigation. For this purpose, they conduct experimental data verification tasks using a set of open source datasets (e.g., Hail-a-Taxii\footnote{http://hailataxii.com/}). They train machine learning models using \attack\ TTPs and use them to identify and detect malware behind cyber incidents from multiple perspectives (e.g., domain addresses, IP addresses, prevalent TTPs, file types, etc.). Based on the achieved results, the authors believe that the trained models are able to predict threat actors and malicious activities with 86\% accuracy.

\subsubsection{Automatic mapping of vulnerability information to adversary techniques, 2022,~\cite{mendsaikhan2020automatic}} 
\begin{figure}
    \centering
    \includegraphics[width=0.7\columnwidth]{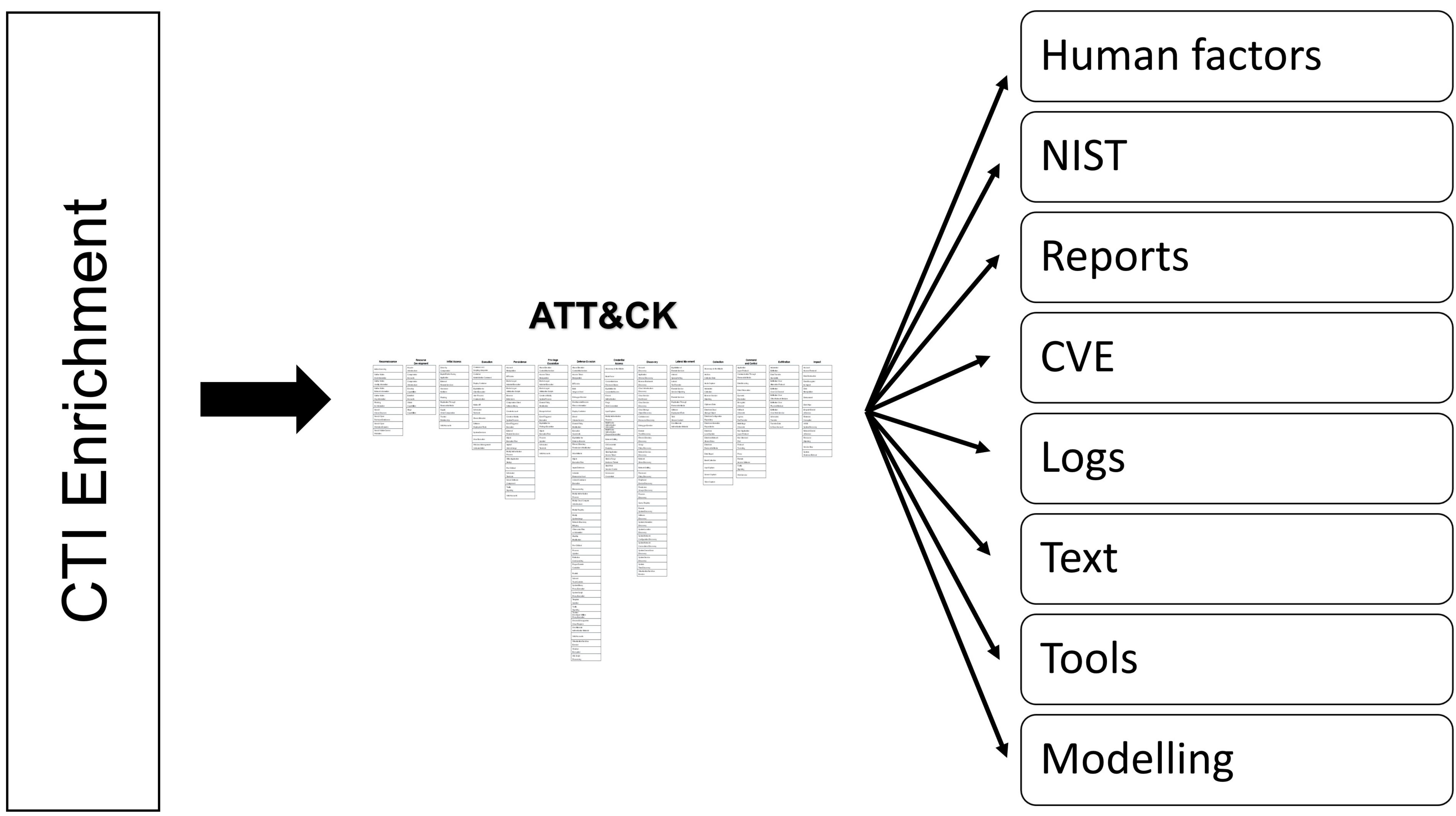}
    \caption{CTI Enrichment with the \attack\ matrix: We identified different sources from the literature, which are combined with the \attack\ matrix to enrich the corporate CTI.} 
    \label{fig:cti_enrichment}
\end{figure}

This research proposes a solution to automate correlating software vulnerabilities to the \mitre\ \attack\ framework using multi-label machine learning classification techniques. For this purpose, they use the vector representation of the vulnerability description (e.g., CVE) as an input to a group of multi-label classification methods. The authors believe that using multi-label classification can improve the process of inferring adversarial techniques because a particular vulnerability can be linked to multiple cyber-threat strategies. For training and evaluation of the classifiers, they use 8,077 examples from a dataset provided by The European Union Agency for Cybersecurity (ENISA). Their results show that the LabelPowerset and Multilayer Perceptron methods performs better than others. 
\\
\\
{\bf Wrap up on {\em CTI enrichment.}} Cyber Threat Intelligence enrichment is a complex process involving several best-practices, methodologies and models to enrich the current state of the CTI. The general idea is rooted on combining information from different sources to generate a more detailed picture of the threat landscape. The literature review has identified several sources that combined with the \attack\ framework can enrich the corporate CTI. Figure~\ref{fig:cti_enrichment} shows the identified sources, i.e., Human factors (\cite{georgiadou2021assessing}), NIST (\cite{kwon2020cyber}), Reports (\cite{legoy2020automated}, \cite{husari2019learning}, \cite{wu2021price}, \cite{ejaz2022visualizing}), CVE (\cite{kuppa2021linking}, \cite{ampel2021linking}), Logs (\cite{choi2020expansion}, \cite{parmar2019use}), Text (\cite{husari2017ttpdrill}, \cite{you2022tim}, \cite{liu2022threat}), and finally, Tools (\cite{stech2016integrating}). A set of papers \cite{elitzur2019attack}, \cite{kim2021automatically}, \cite{nisioti2021data}, \cite{zych2022enhancing} just considers the \attack\ matrix as unique source, while 
\cite{mavroeidis2017cyber} and \cite{maymi2017towards} propose new modelling frameworks to support the CTI enrichment.

\subsection{Defensive Gap Assessment}

{\em Defensive Gap Assessment} is a critical component of the security risk assessment process. Investigating the gaps of the defensive strategies deployed in a corporate ICT infrastructure is of paramount importance---this including the analysis of the tools, procedures and policies to detect and mitigate cyber attacks. The \attack\ framework can support the defensive gap assessment given the huge amount of techniques and procedures it contains: assuming a specific business, we can identify the APTs targeting it, and finally, all the possible tactics, techniques and procedures creating a consistent threat landscape thus enabling the assessment of the current defensive status of the organization.

\subsubsection{Security Assessment Rating Framework for Enterprises using \mitre\ \attack\ Matrix, 2021,~\cite{manocha2021security}} This work introduces a \mitre\ \attack\ based cyber security analytical rating framework enabling organizations to validate their organizational security aspects. The main idea of this research is to create a risk assessment framework that: (i) collects various \mitre\ \attack\ techniques, (ii) assign them severity scores based on their impact on the underlying context and the complexity posed to an attacker, (iii) provide their success/failure states from the collected reports and tests. The outcome will be an organization’s overall risk assessment rating and the respective coverage of the \mitre\ \attack\ matrix per assessment. They believe that this framework helps to reduce the gap between the security operations team and the organization’s strategic development team. The proposed framework has been tested with an example enterprise pen-testing assessment. 

\subsubsection{Cyberattack Models for Ship Equipment Based on the \mitre\ \attack\ Framework, 2022,~\cite{jo2022cyberattack}} This work introduces a \mitre\ \attack\ based cybersecurity analysis approach specifically for ship equipment. Authors investigate and analyze potential cybersecurity threats on ships using \mitre\ \attack\ framework, and present a cyberattack analysis model. For this purpose, they analyze recent ship cybersecurity research projects and provide essential defense recommendations. Using the proposed model, they list ship navigation, communication and control characteristics; outline main threats and vulnerabilities; and provide mitigation strategies. They evaluated the proposed cyberattack analysis model using 4 case studies: case (i) a ship sinking, case (ii) and (iii) wrong routes, case (iv) an attack on the ship control system. They believe that the obtained results can lead them to conduct their next future work. 

\subsubsection{\mitre\ \attack\ Based Evaluation on In-Network Deception Technology for Modernized Electrical Substation Systems, 2022,~\cite{mashima2022mitre}} Authors introduce a design, implementation and deployment of an in-network deception technology for IEC 61850 standard compliant smart substation systems in smart grid called DecIED. Authors main objective is analyzing various deployment strategies of DecIED in order to integrate it into a production environment.  Moreover, they verify DecIED's performance and efficiency against recently emerged high-profile attacks targeting smart grids. Accordingly, they investigate the procedure of using DecIED for mitigating ICS related attacks. Their evaluation is accomplished based on the \mitre\ \attack\ Matrix for industrial control systems which provides corresponding tactics and techniques. They evaluate DecIED's effectiveness against well-known real-world cyberattacks, such as Stuxnet\footnote{https://www.trellix.com/en-sg/security-awareness/ransomware/what-is-stuxnet.html} and Ukraine incidents~\cite{zetter2016inside}, and show its performance and efficiency in such contexts.

\subsubsection{Services for Zero Trust Architectures - A Research Roadmap, 2021,~\cite{bertino2021services}} Authors use information about existing cyber risks as a knowledge-base to propose a Zero Trust Architecture (ZTA) design methodology. Zero Trust Architecture (ZTA) introduces an interesting paradigm for enhancing cybersecurity by supporting well-designed, comprehensive and continues security functions in all organizational levels. ZTA's initial assumption is that no organizational entity is trusted. This paper discusses the key pillars of a ZT model and determines deployment challenges both in research and practice. The authors describe the following pillars for ZT model: (i) collecting and analyzing external and internal threats information, (ii) fundamental security patterns, (iii) ZT security patterns, (iv) identifying and choosing services, technology and process improvements, (v) security service implementation, (vi) security control testing, (vii) performance metrics to measure the effectiveness of the implemented solution. Next, they have elaborated on those pillars and challenges and provided potential solutions. Accordingly, They employ \mitre\ \attack\ framework to design security patterns. 

\subsubsection{Application of Advanced Persistent Threat Actors’ Techniques for Evaluating Defensive Countermeasures, 2020,~\cite{golushko2020application}} This research describes the results of integrating the \mitre\ \attack\ knowledge-base TTPs into the process of defining cybersecurity requirements for information security systems. For this purpose, they follow 3 major phases: (i) preparation stage that defines systems structural and functional features, (ii) threat modeling stage that is the process of identifying potential origins of information security cyber-threats and determining tactics and techniques of the attacker's behavior regarding the structural and functional features, (iii) security controls selection that is the process of selecting protective security metrics.

\subsubsection{A Model and Method of Information System Security Risk Assessment based on \mitre\ \attack\, 2021,~\cite{he2021model}} This research presents Advantageous Tactics, Technologies, and Common Knowledge - Optimized Analytic Hierarchy Process - Bayesian Network (ATT\&CK-OAHP-BN) that is a risk assessment approach based on \mitre\ \attack\ framework. In other words, authors use the \mitre\ \attack\ framework to design and develop a hierarchical risk assessment indicators system. Next, using OAHP algorithm, a weight is assigned to each indicator and its initial risk value is calculated. Finally, they use the Bayesian network inference model to calculate each risk indicator probability to happen and discover the most important risks. Based on their experiments on the use case data of the industrial control CPS, they believe that the proposed model can effectively assign appropriate risk values to security threats, and identify the most important risks to be carefully monitored when security threats happen.
\\\\
{\bf Wrap up on {\em Defensive Gap Assessment.}} Just a minor amount of papers focuses on this use case although the importance of the \attack\ matrix is well recognized in the assessing of the defensive gaps for the ICT infrastructure. Out of the 5 identified papers, we highlight that 3 of them (\cite{jo2022cyberattack},\cite{bertino2021services},\cite{he2021model}) propose new models for gap assessment, one studies the deployment of a rating framework for the security assessment~\cite{manocha2021security}, while the last one takes into account an in-network deception infrastructure and the related assessment for its deployment~\cite{mashima2022mitre}.

\section{Summary and Key findings}
\label{sec:summary-analysis}
This section discusses the major findings of our analysis from different perspectives, such as the application scenario, data sources, adopted techniques, and finally, datasets. 

\subsection{Application scenario}
\label{subsec:application_scenario}

We now focus on the analysis of the {\em application scenarios} as depicted in Table~\ref{tab:application-scenario}. Firstly, out of the considered 56 papers, 16 papers have targeted a particular context/environment/infrastructure to apply their proposed approaches. We identify 5 main categories: (i) Operational Technology, (ii) Defense, (iii) Commercial, (iv) Telecom, (v) Cybersecurity. Each category has its own sub-categories as listed in Table~\ref{tab:application-scenario}. 

Our analysis shows that industrial/utilities applications, such as ICS, power grids, IoT, etc., have attracted most of the research works. We believe that the main reason behind this is rooted in the recent advancements of industrial control systems and the integration with complex ICT infrastructures; indeed, attackers can now perform malicious activities to manipulate the controls of power grids, energy providers, and other critical infrastructure. These activities result in real-world catastrophic physical damage, such as major service outage and various service disruptions. To identify vulnerabilities in industrial systems, security-relevant parts must be understood, potential attacks need to be identified, and all weaknesses that can be exploited should be detected. \mitre\ \attack\ framework is the major knowledge-base for these tasks and accomplishments. Correlating this knowledge-base with other information sources that we will mention in the next sections, is a valuable asset. 

Lastly, we observe that many of the proposed solutions target emerging technologies, e.g., 5G, new generation smartphones, etc., as their {\em application scenarios} highlighting the importance of \mitre\ \attack\ framework for future applications. For example, the recent advances in 5G Core Networks (5GCN) lead to a service-rich technology ecosystem attracting attackers' attention. Hence, numerous 5G network security assessments have been conducted by academics, security groups, and industry suppliers. As another example, social networks attacks, e.g., phishing, is one of the prevalent entry-point toward the compromization of smartphones. Consequently, using continuously updated versions of cybersecurity knowledge-bases like \mitre\ \attack\ framework is critical to resist against emerging attacks. 

\begin{table}
\footnotesize
\caption{Application scenarios.}
\centering
\begin{tabular}{|cll|}
\hline
\multicolumn{3}{|c|}{\textbf{Application Scenarios}}                                                                                                                                      \\ \hline
\multicolumn{1}{|c|}{\multirow{6}{*}{Operational Technology}} & \multicolumn{1}{l|}{Power Grid}          & \cite{hacks2020powerlang}, \cite{hasan2019artificial}, \cite{mashima2022mitre} \\ \cline{2-3} 
\multicolumn{1}{|c|}{}                                        & \multicolumn{1}{l|}{Maritime}            & \cite{jo2022cyberattack}                                                       \\ \cline{2-3} 
\multicolumn{1}{|c|}{}                                        & \multicolumn{1}{l|}{Healthcare}          & \cite{ahmed2022mitre}                                                          \\ \cline{2-3} 
\multicolumn{1}{|c|}{}                                        & \multicolumn{1}{l|}{ICS}                 & \cite{choi2020expansion}, \cite{jadidi2021threat}                              \\ \cline{2-3} 
\multicolumn{1}{|c|}{}                                        & \multicolumn{1}{l|}{IoT}                 & \cite{wu2020grouptracer}, \cite{chierzi2021evolution}                          \\ \cline{2-3} 
\multicolumn{1}{|c|}{}                                        & \multicolumn{1}{l|}{Water System}        & \cite{toker2021mitre}                                                          \\ \hline
\multicolumn{1}{|l|}{Defense}                                 & \multicolumn{1}{l|}{NATO NCI}            & \cite{parmar2019use}                                                           \\ \hline
\multicolumn{1}{|l|}{\multirow{2}{*}{Commercial}}             & \multicolumn{1}{l|}{Finance}             & \cite{xiong2022cyber}                                                          \\ \cline{2-3} 
\multicolumn{1}{|l|}{}                                        & \multicolumn{1}{l|}{E-Commerce}          & \cite{wu2021price}                                                             \\ \hline
\multicolumn{1}{|l|}{\multirow{2}{*}{Telecom}}                & \multicolumn{1}{l|}{5G}                  & \cite{pell2021towards}                                                         \\ \cline{2-3} 
\multicolumn{1}{|l|}{}                                        & \multicolumn{1}{l|}{Cellphone}           & \cite{kadhimi2022}                                                             \\ \hline
\multicolumn{1}{|l|}{Cybersecurity}                           & \multicolumn{1}{l|}{Malware   Detection} & \cite{fairbanks2021identifying}                                                \\ \hline
\end{tabular}
\label{tab:application-scenario}
\end{table}

\subsection{Data sources}
\label{subsec:data_sources}

In this section, we provide a categorization of the (input) data taken into account by the papers considered in this work. 
The considered data is adopted for different purposes such as designing, developping and validating the adopted methodologies. Table~\ref{fig:input-table} summarizes the input data used in the considered papers. While the \mitre\ \attack\ framework constitutes a common knowledge for all the considered papers to design, develop, extend, enhance or validate their proposed solutions, other data sources are: (i) \mitre\ CVE, (ii) \mitre\ CAPEC, (iii) NIST, (iv) event logs/audits, (v) security reports. 

\mitre\ CVE, \mitre\ CAPEC and NIST are mainly used as datasets to be correlated with the \mitre\ ATT\&CK framework. As previously discussed, \mitre\ CVE reports the vulnerabilities discovered in a wide range of software and hardware, thus linking it with the \mitre\ \attack\ framework generates an improved ground knowledge of tactics, techniques and associated vulnerabilities. \mitre\ CAPEC focuses on the application security and enumerates exploits against vulnerable systems. It is mainly used for application threat modeling and developer training and education. Hence, it can extend \mitre\ \attack\ knowledge-base with these information. NIST provides a wide range of defense mechanisms in different domains: identification, protection, detection and recovering. Adding it to the \mitre\ \attack\ framework results a more enhanced knowledge-base. 

\begin{table}
\footnotesize
    \centering
    \caption{Data sources.}
    \begin{tabular}{|l|l|}
    \hline
        \textbf{Input} & \textbf{Papers} \\ \hline \hline
        
        \textbf{MITRE CVE} &  \cite{kuppa2021linking}, 
 \cite{ampel2021linking}, 
 \cite{nisioti2021game}, 
 \cite{mendsaikhan2020automatic} \\ \hline
        \textbf{MITRE CAPEC} &  \cite{husari2017ttpdrill}, 
 \cite{farooq2018optimal}, 
 \cite{mendsaikhan2020automatic} \\ \hline
        \textbf{NIST} & \cite{kwon2020cyber} \\ \hline
        \textbf{Logs/Audits} &  \cite{milajerdi2019holmes}, 
 \cite{hassan2020tactical}, 
 \cite{choi2020expansion}, 
 \cite{kim2021automatically},
 \cite{wu2020grouptracer}, 
 \cite{toker2021mitre}, 
 \cite{huang2021open}, 
 \cite{takey2021real} \\ \hline
        \textbf{Security Reports} &  \begin{tabular}[c]{@{}l@{}} \cite{husari2017ttpdrill}, 
 \cite{noor2019machine}, 
 \cite{stech2016integrating},
 \cite{oosthoek2019sok},
 \cite{georgiadou2021assessing}, 
 \cite{munaiah2019characterizing}, 
 \cite{legoy2020automated}, 
 \cite{husari2019learning}, \\
 \cite{elitzur2019attack}, 
 \cite{kuppa2021linking}, 
 \cite{kim2021automatically}, 
 \cite{karuna2021automating}, 
 \cite{jadidi2021threat}, 
 \cite{manocha2021security}, 
 \cite{wu2021price}, 
 \cite{you2022tim}, \\
 \cite{jo2022cyberattack}, 
 \cite{ahmed2022mitre}, 
 \cite{nisioti2021game}, 
 \cite{pell2021towards}, 
 \cite{kadhimi2022}, 
 \cite{liu2022threat}, 
 \cite{chierzi2021evolution}, 
 \cite{zych2022enhancing}, 
 \cite{ejaz2022visualizing}\end{tabular}\\ \hline
    \end{tabular}
    \label{fig:input-table}
\end{table}

Event logs and audit data, e.g., {\em Linux auditd} and {\em Windows ETW data}, along with security reports have been used as major sources to automatically discover APTs. The goal is mapping activities found in logs and alerts found in security monitoring systems to the APT kill chain. This provides deep visibility into the attacker TTPs and facilitate threat investigation. However, there are still challenges questioning their usefulness in practice, e.g., validation and investigation, long-term log storage, automated information extraction and analysis, etc. Machine learning and NLP techniques have been widely used to address these challenges and improve the usage of these information sources in real environments and applications. 

\subsection{Adopted Techniques}
\label{subsec:adopted-techniques}

Each considered paper employs or integrates different techniques in order to build and design their methodologies while having \mitre\ \attack\ as their main building block. Table~\ref{tab:aod-table} lists the adopted techniques and tools for the considered papers. Techniques, such as information correlation, ontology engineering, Natural Language Processing (NLP), machine learning, Domain Specific Language (DSL) design and simulating test-beds, have been considered. In addition, some papers integrate security tools (e.g., Intrusion Detection Systems (IDS), Security Information and Security Information and Event Management (SIEM)) to their proposed methodologies. The main objective of using information correlation and ontology engineering is to enhance/improve/extend the existing knowledge-base, \mitre\ \attack\ in particular. Table~\ref{fig:input-table} shows the list of inputs to these techniques in our list of papers. NLP and Machine learning techniques help to automate the knowledge extraction process and split data into well-defined and more accurate classes/clusters for more detailed analysis (e.g., identifying TTPs based on low-level threat artifacts observed in the underlying environment). Among all the adopted techniques, machine learning is the most diversified domain including a large list of models/sub-techniques, such as hierarchical clustering, decision trees, Support Vector Machines (SVM), neural networks, Hidden Markov Models (HMM), etc. Some papers use more than one model to compare their accuracy (experimental analysis) or for complementary purposes. However, machine learning based analysis should be performed through an optimal workflow to ensure well prepared training/testing datasets and fine-tuned models (predictor, classifier, etc.) for subsequent analysis. This help to address the high volume of false positives and filtering out irrelevant results. 

\begin{table}
\footnotesize
\centering
\caption{Adopted techniques.}
\label{tab:aod-table}
\begin{tabular}{|l|l|}
\hline
\textbf{Adopted   Technique} & \textbf{Papers} \\ \hline
Information   Correlation & \begin{tabular}[c]{@{}l@{}}\cite{milajerdi2019holmes},   \cite{mavroeidis2017cyber}, \cite{husari2017ttpdrill},   \cite{stech2016integrating}, \\      \cite{farooq2018optimal}, \cite{georgiadou2021assessing},   \cite{kwon2020cyber}, \cite{munaiah2019characterizing}, \cite{straub2020modeling},   \cite{nisioti2021data}, \cite{legoy2020automated}, \cite{husari2019learning},   \cite{parmar2019use}, \cite{berady2021ttp}, \cite{elitzur2019attack},   \cite{choi2020expansion}, \cite{kim2021automatically},   \cite{golushko2020application}, \cite{wu2020grouptracer},   \cite{ampel2021linking}, \cite{jadidi2021threat}, \\      \cite{bertino2021services}, \cite{huang2021open},   \cite{wang2020clustering}, \cite{takey2021real}, \cite{wu2021price},   \cite{jo2022cyberattack}, \cite{mashima2022mitre}, \cite{nisioti2021game},   \\      \cite{pell2021towards}, \cite{kadhimi2022}, \cite{chierzi2021evolution},   \cite{fairbanks2021identifying}, \\      \cite{he2021model}, \cite{zych2022enhancing},\\      \cite{huang2021open}, \cite{wang2020clustering}, \cite{takey2021real},   \cite{fairbanks2021identifying}, \\      \cite{he2021model},\end{tabular} \\ \hline
Ontology & \cite{mavroeidis2017cyber},   \cite{husari2017ttpdrill}, \cite{husari2019learning}, \\ \hline
NLP & \begin{tabular}[c]{@{}l@{}}\cite{husari2017ttpdrill},   \cite{legoy2020automated}, \cite{husari2019learning},   \cite{karuna2021automating}, \\      \cite{wu2021price}, \cite{you2022tim}, \cite{liu2022threat}\end{tabular} \\ \hline
ML & \begin{tabular}[c]{@{}l@{}}\cite{rawan2020},   \cite{noor2019machine}, \cite{farooq2018optimal}, \cite{nisioti2021data},   \\      \cite{legoy2020automated}, \cite{choi2021probabilistic},   \cite{hasan2019artificial}, \\      \cite{elitzur2019attack}, \cite{kuppa2021linking}, \cite{karuna2021automating},   \cite{wu2020grouptracer}, \\      \cite{toker2021mitre}, \cite{huang2021open}, \cite{wang2020clustering},   \cite{takey2021real}, \cite{liu2022threat}, \cite{fairbanks2021identifying},   \cite{he2021model}, \cite{ejaz2022visualizing},\end{tabular} \\ \hline
DSL & \begin{tabular}[c]{@{}l@{}}\cite{xiong2022cyber},   \cite{hacks2020powerlang}, \cite{maymi2017towards},   \cite{karuna2021automating},\\      \cite{hassan2020tactical},\end{tabular} \\ \hline
Simulation   test-bed & \cite{applebaum2017analysis} \\ \hline
IDS/SIEM & \cite{toker2021mitre} \\ \hline
\end{tabular}
\end{table}

\subsection{Datasets}
\label{subsec:datasets}

Table~\ref{tab:dataset} shows the list of publicly available datasets referenced by the considered paper in this work. These datasets can be useful to extend/enhance existing approaches toward their future objectives. We classify these datasets into 4 major groups: 

\begin{itemize}
  \item[-] CTI repositories 
  \item[-] Malware databases
  \item[-] TTPs
  \item[-] Event logs
\end{itemize}

Some of these datasets have been prepared specifically for Industrial Control Systems cyber-threat evaluation (e.g., SWaT dataset). Some others provide raw texts as threat and vulnerability reports to be analyzed by NLP techniques. Overall, we believe that this list can be significantly useful for future work in the area as summarized in the next section. 

\begin{table}
\footnotesize
\caption{List of considered datasets.}
\label{tab:dataset}
\begin{tabular}{|l|l|l|l}
\cline{1-3}
\textbf{Paper} & \textbf{Dataset} & \textbf{Description} &  \\ \cline{1-3}
\cite{ejaz2022visualizing} & Hail-a-Taxii dataset\footnote{http://hailataxii.com/} & Open source CTI repository &  \\ \cline{1-3}
\cite{ejaz2022visualizing} & TTP Dataset \cite{noor2019machine} & Open source dataset adopted to train machine learning models. &  \\ \cline{1-3}
\cite{milajerdi2019holmes} & DARPA Transparent Computing\footnote{https://github.com/darpa-i2o/Transparent-Computing} & \begin{tabular}[c]{@{}l@{}} An expert red-team simulating multiple attacks on \\a segmented computer network \end{tabular} &  \\ \cline{1-3}
\cite{husari2019learning} & Symantec\footnote{https://www.broadcom.com/support/security-center?} and Fireeye CTI blogs\footnote{https://www.mandiant.com/advantage/threat-intelligence} & Symantec and Firewall blogs &  \\ \cline{1-3}
\cite{jadidi2021threat} & A six-stage Secure Water Treatment (SWaT) dataset\footnote{https://itrust.sutd.edu.sg/itrust-labs\_datasets/dataset\_info/} & \begin{tabular}[c]{@{}l@{}} A small-size industrial water analysis procedure which was\\ generated by iTrust Cyber Security Research Centre. \end{tabular} &  \\ \cline{1-3}
\cite{berady2021ttp} & APT29 Activity From the ATT\&CK Evaluations\footnote{https://github.com/OTRF/detection-hackathon-apt29} & \begin{tabular}[c]{@{}l@{}} Includes the generated logs by re-running both parts of an attack \\ scenario \end{tabular} &  \\ \cline{1-3}
\cite{wang2020clustering} & ATT\&CK for Industrial Control Systems: Groups\footnote{https://attack.mitre.org/groups/} & It provides an index of APT groups &  \\ \cline{1-3}
\cite{huang2021open} & MalShare\footnote{https://malshare.com/} & Malware database &  \\ \cline{1-3}
\cite{ampel2021linking} & BRON knowledge graph dataset \cite{hemberg2020bron} & \begin{tabular}[c]{@{}l@{}} Linking Attack Tactics, Techniques, and Patterns with\\ defensive weaknesses, vulnerabilities and affected \\environment configurations \end{tabular} &  \\ \cline{1-3}
\cite{karuna2021automating} & Operationally Transparent Cyber (OpTC) evaluation set\footnote{https://github.com/FiveDirections/OpTC-data} & \begin{tabular}[c]{@{}l@{}} Operationally Transparent Cyber (OpTC) was a technology\\ transition pilot study funded under Boston Fusion Corp.'s \\Cyber APT Scenarios for EnterpriseSystems (CASES) project \end{tabular} &  \\ \cline{1-3}
\cite{karuna2021automating} & Caldera\footnote{https://caldera.readthedocs.io/en/latest/} & \begin{tabular}[c]{@{}l@{}} A security platform to run autonomous breach-and-simulation\\ experiments.  \end{tabular} &  \\ \cline{1-3}
\cite{choi2020expansion} & HAI testbed dataset \cite{shin2019implementation} & \begin{tabular}[c]{@{}l@{}} The hardware-in-the-loop (HIL)-based augmented ICS (HAI)\\ testbed and dataset \end{tabular}&  \\ \cline{1-3}
\cite{kuppa2021linking} & APT reports that are published from 2008 to 2019\footnote{https://github.com/CyberMonitor/APT\_CyberCriminal\_Campagin\_Collections} & APT \& cybercriminals campaign collection &  \\ \cline{1-3}
\cite{kuppa2021linking} & Zero-day exploits observed by google project zero\footnote{https://googleprojectzero.blogspot.com/p/0day.html} & This data is collected from a range of public sources. &  \\ \cline{1-3}
\cite{kuppa2021linking} & 63720 vulnerability reports\footnote{https://www.symantec.com/security-center/vulnerabilities} & Vulnerability reports from Symantec &  \\ \cline{1-3}
\cite{kuppa2021linking} & 37000 threat reports\footnote{https://www.symantec.com/security-center/a-z} & Threat reports from Symantec &  \\ \cline{1-3}
\cite{oosthoek2019sok} & Malpedia \cite{plohmann2017malpedia} & A Collaborative effort to inventorize the malware landscape &  \\ \cline{1-3}
\cite{mendsaikhan2020automatic} & State of vulnerabilities 2018/2019\footnote{https://github.com/enisaeu/vuln-report/} & \begin{tabular}[c]{@{}l@{}} European Union Agency for Cybersecurity (ENISA),\\ Tech Report, 2019 \end{tabular}&  \\ \cline{1-3}
\end{tabular}
\end{table}

\section{Open issues and research directions}
\label{sec:open-issues-and-research-directions}

In this section, we describe current research challenges and possible future directions. Our analysis is based on the  previously introduced categorization in Section~\ref{sec:use_cases_mapping}.

\subsection{Challenges and future directions in behavioural analytic}
\begin{itemize}
  \item[-] \textbf{Data source analysis:} data source is a critical factor biasing the behavioral analytic. Current challenges on data source analysis are: (i) current approaches considers each endpoint independently and cross-machine provenance has not been fully analyzed yet, (ii) most of the existing implementations and experiments are offline, while real-time analysis has still challenges, such as integrating new data structures and threat scoring algorithms, (iii) adaptive attacks employ tactics in an unpredictable way to bypass detection algorithms. Advanced scoring algorithms are required to address this challenge, (iv) real APT behaviors instead of simulated ones are required to conduct comprehensive analysis and experiments, (v) some attacks do not trigger alerts, which in turn shoul be generated by underlying EDR systems.
  
  \item[-] \textbf{Learning techniques associations:} reliable construction of technique associations provides insights and knowledge to predict 0-day attack techniques. Future research directions in this field are: (i) preparing a ground truth attack dataset in order to enhance the associations quality and comprehensiveness, (ii) investigating more machine learning approaches (e.g., hierarchical clustering) to investigate probabilistic technique associations to reveal pre-conditions and post-conditions relationships.

  \item[-] \textbf{Similarity analysis:} this technique can be used to capture similarity degrees of adversary behaviors based on various semantic links among APT groups. The results can be used as threat intelligence information for advanced threat analysis. Future directions can be: (i) analyzing the similarity of attackers activities and behavior to learn attackers habits and enrich the ground-truth of attack patterns, (ii) attacks have time-lines, and the time dimension can be added to analyze the attackers' mindset.

\item[-] \textbf{Characterizing attackers behaviors:} analyzing various datasets, in particular those collected from penetration testing competitions, can provide valuable insights in order to discover and exploit software systems weaknesses on a typical way. The results can improve security experts knowledge toward the goal of designing and developing secure software systems.  

\item[-] \textbf{Modeling attacks and defenses:} multiple propositions for security attacks and defense modeling have been proposed, e.g., BACCER~\cite{straub2020modeling} in Section~\ref{sec:use_cases_mapping}. Continued refinement of these modeling systems and using them within various use case-targeted demonstrations can be a suitable future direction. Moreover, evaluating the modeling systems based on multiple real scenarios and analyzing the results in order to apply to broader cybersecurity applications can be another future direction.

\item[-] \textbf{Case Study - Industrial Control Systems:} ICS threat hunting frameworks have been extensively studied and investigated. The fist phase of the threat hunting process is the hypothesis proposition that is defining a list of potential threats to be verified and investigated using CTI information. Automating hypothesis verification (quality and reliability) and threat intelligence generation tasks can improve response issuing time. Consequently, the future directions can be: (i) improving initial hypothesis reliability, (ii) automating hypothesis verification and threat intelligence generation using machine learning techniques. 

\item[-] \textbf{Real-time multi-step cyberattack detection:} APT attacks are often categorized as multistage attacks. During such attacks, attackers drop malware on the victim systems and try to distribute and transfer it to the whole organization. Future directions might involve (i) developing more advanced and intelligent techniques for detecting multistage attacks, (ii) early multistage attacks detection that needs gaining knowledge about the most recent adopted techniques by adversaries and the end-to-end procedure of such attacks.

\item[-] \textbf{Risk assessment}: MITRE \attack\ based risk assessment has been studied by researchers in the recent years as it reveals information and knowledge about attackers' interests, motivations, abilities, etc. Integrating organization-specific information is another missing contribution in the current research, e.g., collecting information from/about honeypots and indicators of compromises (IoCs), etc. 

\item[-] \textbf{Case Study - Internet of Things (IoT):} researchers have studied the advancement of techniques and strategies utlized by IoT malware by leveraging MITRE ATT\&CK matrix in order to analyze and characterize (e.g., the pre- and post- exploitation aspects of an attack) current IoT related cyber attacks. Developing an efficient automated IoT malware investigation along with expanding the research to cover emerging threats can be important research directions.  

\end{itemize}

\subsection{CTI enrichment challenges and future directions}
\begin{itemize}

\item[-] \textbf{Cyber threat intelligence ontology:} representing and sharing cybersecurity analytical data and information in an optimized fashion require a unified format and a set of standard protocols between researchers other entities. A solution proposed by researchers is the use of ontologies. Although various ontologies have been proposed in the literature, a well-designed multi-layered CTI ontology is still an important need to be considered as a research direction.

\item[-] \textbf{Automatic and accurate extraction of CTI data from unstructured texts:} automated extraction and analysis of CTI information from various unstructured and free-style reports and texts have attracted researchers to propose appropriate solutions. These research directions are still valid: (i) using popular NLP solutions (e.g., Google NLP and Azure Language Studio) to extract and analyze CTIs from cybersecurity reports and comparing the results accuracy and efficiency, (ii) extending CTI reports repositories to cover various cyberthreat threat source and languages, (iii) constructing a TTP graph for enhanced extraction, investigation and prediction of emerging TTPs.

\item[-] \textbf{Learning APT chains from cyber threat intelligence:} using machine learning techniques to learn temporal relations between malicious artifacts in order to provide insights for threat analysts of the malicious activities happening on the underlying environment. 

\item[-] \textbf{Attack hypothesis generator:} leveraging the advantages of CTI knowledge and insights in order to generate a set of hypotheses for the underlying environment facilitates forensic analysts analytical tasks. Future research directions might consider including features matching with domain expert knowledge, such as target operating system, kill chain steps and information sources, can improve attack hypothesis generator's accuracy. 

\item[-] \textbf{Expanding MITRE ATT\&CK framework:} MITRE ATT\&CK framework can be extended to include more information sources, such as CVE, CWE, CAPEC, NIST, etc. In addition, future research direction can be: (i) conducting experiments to compare the expanded framework performance on bench-marking datasets in other fields, (ii) examining the security strength of the extended framework under adversarial settings and threat models, (iii)  investigating more refined textual data representation techniques (e.g., novel  word embedding strategies, synonym/homonym generation, POS tagging) to improve the features of input texts.
\end{itemize}

\subsection{Adversary emulation / Red teaming}
\begin{itemize}
\item[-] \textbf{Threat modeling:} this concept is used to identify the list of existing assets and resources within an environment and corresponding potential threats. Threat modeling is used to validate security status of an existing system or to develop a new system with the security-by-design strategy. An important future direction is to calculate and assign probability scores to cyberattack steps/stages to make the modeling and simulation results more and more realistic.

\item[-] \textbf{Adversary emulation:}  adversary emulation provides a solid method to verify a system resilience against various advances cyberattacks. However, automating this task still has several technical challenges, such as techniques encoding, and the coordination of their deployment.

\item[-] \textbf{Advanced persistent graphs for threat hunting:} such graphs represent the attack patterns of the adversary through the network during the persistent attack campaign. As future directions, we refer to all the techniques involving graph similarity computation in order to generate metrics to reliably estimate the effectiveness of the attack and the defense.

\item[-] \textbf{Case study - dynamic threat modeling in 5G Core Networks (5GCN):} as the APTs continue to expand and cover emerging technologies, a potential target for threat actors is 5GCN. Hence, several security researchers, analysts and industry suppliers have proposed different 5GCN security assessment approaches and solutions. Future research directions can be: (i) generating datasets that include identified 5GCN attack scenarios and conducting advanced security analysis and assessments. (ii) analytical study to prepare a comprehensive list of requirements for 5GCN adversarial techniques detection and mitigation, (iii) proposing an appropriate risk assessment approach to cover emerging 5GCN TTPs and provide complementary 5GCN risk evaluation recommendations, (iv) proposing an analytical framework based on game-theory to model adversary and defender's activities within 5GCN infrastructures.

\end{itemize}

\subsection{Defensive Gap Assessment}
\begin{itemize}

\item[-] \textbf{Zero Trust Architectures (ZTA):} the main objective of a ZTA model is protecting organizations assets, such as data, systems and services, against attackers. Hence, there are a list of challenges to address as future research directions: (i) updating current digital infrastructure to be compatible with ZTA is costly and introduces various technical and knowledge-wide difficulties, (ii) compatibility and integration of ZTA with cloud-based platforms has still unsolved challenges, (iii) integrating ZTA to emerging technologies (e.g., 5G/6G networks) will oppose novel challenges.

\item[-] \textbf{Case study - cyberattack models for ship equipment:} upgrading ship technology and integrating high-end IT and communication technologies to this industry introduces novel security challenges. Future research directions might include: (i) design and develop risk management approaches compatible with the ship industry, (ii) developing a honeypot system to identify malicious activities, potential attack and help to improve threat detection accuracy by reducing false alarm rate, 
(iii) leveraging AI benefits and integrating them to ship cyber-threat detection and response technology, (iv) using maritime CTI information for analyzing, classification and verification of maritime cyber-threat analysis, (v) integrating the digital forensic technology to maritime for analyzing cyber incidents of onboard systems. 

\end{itemize}

\section{Conclusion}
\label{sec:conclusion}

This paper introduces a detailed analysis of the \mitre\ \attack\ framework while considering the state of the art contributions in the field. We have provided a description of the \mitre\ \attack\ framework while comparing it with other threat intelligence frameworks such as STRIDE and the Cyber kill chain. We classified all the works considered in our analysis by use cases, i.e., behavioural analytic, red teaming, CTI enrichment, and finally, defensive gap assessment. Moreover, we categorized the considered works as per application scenarios, data sources, adopted techniques, and datasets. Finally, we discussed open issues and potential research directions, while drawing some final remarks.

\section*{Acknowledgements}
This publication was made possible by the awards NPRP12S-0125-190013 and GRSA7-1-0516-20061 from Qatar National Research Fund (a member of Qatar Foundation). The contents herein are solely the responsibility of the author[s].

\bibliographystyle{IEEEtran}
\bibliography{main}

\end{document}